\documentclass[twocolumn]{aastex631}

\usepackage{graphicx} 
\usepackage{amsmath}
\usepackage{color}
\usepackage{mathtools}
\usepackage{hyperref}

\newcommand {\etal}{et\,al.\,}

\begin{document}

\title{CEERS Key Paper III: The Diversity of Galaxy Structure and Morphology at \boldmath{$z=3-9$} with JWST}

\received{October 27, 2022}
\revised{November 18, 2022}
\accepted{November 28, 2022}

\submitjournal{ApJL}

\shorttitle{CEERS Structure \& Morphology at $z>3$}
\shortauthors{Kartaltepe et al.}

\email{jeyhan@astro.rit.edu}

\author[0000-0001-9187-3605]{Jeyhan S. Kartaltepe}
\affil{Laboratory for Multiwavelength Astrophysics, School of Physics and Astronomy, Rochester Institute of Technology, 84 Lomb Memorial Drive, Rochester, NY 14623, USA}

\author[0000-0002-8018-3219]{Caitlin Rose}
\affil{Laboratory for Multiwavelength Astrophysics, School of Physics and Astronomy, Rochester Institute of Technology, 84 Lomb Memorial Drive, Rochester, NY 14623, USA}

\author[0000-0002-8163-0172]{Brittany N. Vanderhoof}
\affil{Laboratory for Multiwavelength Astrophysics, School of Physics and Astronomy, Rochester Institute of Technology, 84 Lomb Memorial Drive, Rochester, NY 14623, USA}

\author[0000-0001-8688-2443]{Elizabeth J.\ McGrath}
\affiliation{Department of Physics and Astronomy, Colby College, Waterville, ME 04901, USA}

\author[0000-0001-6820-0015]{Luca Costantin}
\affiliation{Centro de Astrobiolog\'{\i}a (CAB), CSIC-INTA, Ctra. de Ajalvir km 4, Torrej\'on de Ardoz, E-28850, Madrid, Spain}

\author[0000-0002-1803-794X]{Isabella G. Cox}
\affiliation{Laboratory for Multiwavelength Astrophysics, School of Physics and Astronomy, Rochester Institute of Technology, 84 Lomb Memorial Drive, Rochester, NY 14623, USA}

\author[0000-0003-3466-035X]{L. Y. Aaron\ Yung}
\altaffiliation{NASA Postdoctoral Fellow}
\affiliation{Astrophysics Science Division, NASA Goddard Space Flight Center, 8800 Greenbelt Rd, Greenbelt, MD 20771, USA}

\author[0000-0002-8360-3880]{Dale D. Kocevski}
\affiliation{Department of Physics and Astronomy, Colby College, Waterville, ME 04901, USA}

\author[0000-0003-3735-1931]{Stijn Wuyts}
\affiliation{Department of Physics, University of Bath, Claverton Down, Bath BA2 7AY, UK}

\author[0000-0001-7113-2738]{Henry C. Ferguson}
\affiliation{Space Telescope Science Institute, 3700 San Martin Drive, Baltimore, MD 21218, USA}

\author[0000-0002-9921-9218]{Micaela B. Bagley}
\affiliation{Department of Astronomy, The University of Texas at Austin, Austin, TX, USA}

\author[0000-0001-8519-1130]{Steven L. Finkelstein}
\affiliation{Department of Astronomy, The University of Texas at Austin, Austin, TX, USA}

\author[0000-0001-5758-1000]{Ricardo O. Amor\'{i}n}
\affiliation{Instituto de Investigaci\'{o}n Multidisciplinar en Ciencia y Tecnolog\'{i}a, Universidad de La Serena, Raul Bitr\'{a}n 1305, La Serena 2204000, Chile}
\affiliation{Departamento de Astronom\'{i}a, Universidad de La Serena, Av. Juan Cisternas 1200 Norte, La Serena 1720236, Chile}

\author[0000-0001-8085-5890]{Brett H.~Andrews}
\affiliation{Department of Physics and Astronomy, University of Pittsburgh, Pittsburgh, PA 15260, USA}
\affiliation{Pittsburgh Particle Physics, Astrophysics, and Cosmology Center (PITT PACC), University of Pittsburgh, Pittsburgh, PA 15260, USA}

\author[0000-0002-7959-8783]{Pablo Arrabal Haro}
\affiliation{NSF's National Optical-Infrared Astronomy Research Laboratory, 950 N. Cherry Ave., Tucson, AZ 85719, USA}

\author[0000-0001-8534-7502]{Bren E. Backhaus}
\affiliation{Department of Physics, 196 Auditorium Road, Unit 3046, University of Connecticut, Storrs, CT 06269, USA}

\author[0000-0002-2517-6446]{Peter Behroozi}
\affiliation{Department of Astronomy and Steward Observatory, University of Arizona, Tucson, AZ 85721, USA}
\affiliation{Division of Science, National Astronomical Observatory of Japan, 2-21-1 Osawa, Mitaka, Tokyo 181-8588, Japan}

\author[0000-0003-0492-4924]{Laura Bisigello}
\affiliation{Dipartimento di Fisica e Astronomia ``G.Galilei", Universit\'a di Padova, Via Marzolo 8, I-35131 Padova, Italy}
\affiliation{INAF--Osservatorio Astronomico di Padova, Vicolo dell'Osservatorio 5, I-35122, Padova, Italy}

\author[0000-0003-2536-1614]{Antonello Calabr{\`o}}
\affiliation{INAF - Osservatorio Astronomico di Roma, via di Frascati 33, 00078 Monte Porzio Catone, Italy}

\author[0000-0002-0930-6466]{Caitlin M. Casey}
\affiliation{Department of Astronomy, The University of Texas at Austin, Austin, TX, USA}

\author[0000-0002-4343-0479]{Rosemary T. Coogan}
\affiliation{CEA, IRFU, DAp, AIM, Universit\'{e} Paris-Saclay, Universit\'{e} Paris Cit\'{e}, Sorbonne Paris Cit\'{e}, CNRS, 91191 Gif-sur-Yvette, France}

\author[0000-0003-1371-6019]{M. C. Cooper}
\affiliation{Department of Physics \& Astronomy, University of California, Irvine, 4129 Reines Hall, Irvine, CA 92697, USA}

\author[0000-0002-5009-512X]{Darren Croton}
\affiliation{Centre for Astrophysics \& Supercomputing, Swinburne University of Technology, Hawthorn, VIC 3122, Australia}
\affiliation{ARC Centre of Excellence for All Sky Astrophysics in 3 Dimensions (ASTRO 3D)}

\author[0000-0002-6219-5558]{Alexander de la Vega}
\affiliation{Department of Physics and Astronomy, University of California, Riverside, CA 92521, USA}

\author[0000-0001-5414-5131]{Mark Dickinson}
\affiliation{NSF's National Optical-Infrared Astronomy Research Laboratory, 950 N. Cherry Ave., Tucson, AZ 85719, USA}

\author[0000-0003-3820-2823]{Adriano Fontana}
\affiliation{INAF - Osservatorio Astronomico di Roma, via di Frascati 33, 00078 Monte Porzio Catone, Italy}

\author[0000-0002-3560-8599]{Maximilien Franco}
\affiliation{Department of Astronomy, The University of Texas at Austin, Austin, TX, USA}

\author[0000-0002-5688-0663]{Andrea Grazian}
\affiliation{INAF--Osservatorio Astronomico di Padova, Vicolo dell'Osservatorio 5, I-35122, Padova, Italy}

\author[0000-0001-9440-8872]{Norman A. Grogin}
\affiliation{Space Telescope Science Institute, 3700 San Martin Drive, Baltimore, MD 21218, USA}

\author[0000-0001-6145-5090]{Nimish P. Hathi}
\affiliation{Space Telescope Science Institute, 3700 San Martin Drive, Baltimore, MD 21218, USA}

\author[0000-0002-4884-6756]{Benne W. Holwerda}
\affil{Physics \& Astronomy Department, University of Louisville, 40292 KY, Louisville, USA}

\author[0000-0002-1416-8483]{Marc Huertas-Company}
\affil{Instituto de Astrof\'isica de Canarias, La Laguna, Tenerife, Spain}
\affil{Universidad de la Laguna, La Laguna, Tenerife, Spain}
\affil{Universit\'e Paris-Cit\'e, LERMA - Observatoire de Paris, PSL, Paris, France}

\author[0000-0001-9298-3523]{Kartheik G. Iyer}
\affiliation{Dunlap Institute for Astronomy \& Astrophysics, University of Toronto, Toronto, ON M5S 3H4, Canada}

\author[0000-0002-1590-0568]{Shardha Jogee}
\affiliation{Department of Astronomy, The University of Texas at Austin, Austin, TX, USA}

\author[0000-0003-1187-4240]{Intae Jung}
\affiliation{Space Telescope Science Institute, 3700 San Martin Drive, Baltimore, MD 21218, USA}

\author[0000-0001-8152-3943]{Lisa J. Kewley}
\affiliation{Center for Astrophysics, Harvard \& Smithsonian, 60 Garden Street, Cambridge, MA 02138, USA}

\author[0000-0002-5537-8110]{Allison Kirkpatrick}
\affiliation{Department of Physics and Astronomy, University of Kansas, Lawrence, KS 66045, USA}

\author[0000-0002-6610-2048]{Anton M. Koekemoer}
\affiliation{Space Telescope Science Institute, 3700 San Martin Drive, Baltimore, MD 21218, USA}

\author{James Liu}
\affiliation{Laboratory for Multiwavelength Astrophysics, School of Physics and Astronomy, Rochester Institute of Technology, 84 Lomb Memorial Drive, Rochester, NY 14623, USA}

\author[0000-0003-3130-5643]{Jennifer M. Lotz}
\affiliation{Gemini Observatory/NSF's National Optical-Infrared Astronomy Research Laboratory, 950 N. Cherry Ave., Tucson, AZ 85719, USA}

\author[0000-0003-1581-7825]{Ray A. Lucas}
\affiliation{Space Telescope Science Institute, 3700 San Martin Drive, Baltimore, MD 21218, USA}

\author[0000-0001-8684-2222]{Jeffrey A.\ Newman}
\affiliation{Department of Physics and Astronomy and PITT PACC, University of Pittsburgh, Pittsburgh, PA 15260, USA}

\author[0000-0003-4196-0617]{Camilla Pacifici}
\affiliation{Space Telescope Science Institute, 3700 San Martin Drive, Baltimore, MD 21218, USA}

\author[0000-0002-2499-9205]{Viraj Pandya}
\altaffiliation{Hubble Fellow}
\affiliation{Columbia Astrophysics Laboratory, Columbia University, 550 West 120th Street, New York, NY 10027, USA}

\author[0000-0001-7503-8482]{Casey Papovich}
\affiliation{Department of Physics and Astronomy, Texas A\&M University, College Station, TX, 77843-4242 USA}
\affiliation{George P.\ and Cynthia Woods Mitchell Institute for Fundamental Physics and Astronomy, Texas A\&M University, College Station, TX, 77843-4242 USA}

\author[0000-0001-8940-6768]{Laura Pentericci}
\affiliation{INAF - Osservatorio Astronomico di Roma, via di Frascati 33, 00078 Monte Porzio Catone, Italy}

\author[0000-0003-4528-5639]{Pablo G. P\'erez-Gonz\'alez}
\affiliation{Centro de Astrobiolog\'{\i}a (CAB), CSIC-INTA, Ctra. de Ajalvir km 4, Torrej\'on de Ardoz, E-28850, Madrid, Spain}

\author{Jayse Petersen}
\affiliation{Laboratory for Multiwavelength Astrophysics, School of Physics and Astronomy, Rochester Institute of Technology, 84 Lomb Memorial Drive, Rochester, NY 14623, USA}

\author[0000-0003-3382-5941]{Nor Pirzkal}
\affiliation{Space Telescope Science Institute, 3700 San Martin Drive, Baltimore, MD 21218, USA}

\author[0000-0002-9946-4731]{Marc Rafelski}
\affiliation{Space Telescope Science Institute, 3700 San Martin Drive, Baltimore, MD 21218, USA}
\affiliation{Department of Physics and Astronomy, Johns Hopkins University, Baltimore, MD 21218, USA}

\author[0000-0002-5269-6527]{Swara Ravindranath}
\affiliation{Space Telescope Science Institute, 3700 San Martin Drive, Baltimore, MD 21218, USA}

\author[0000-0002-6386-7299]{Raymond C. Simons}
\affiliation{Space Telescope Science Institute, 3700 San Martin Drive, Baltimore, MD 21218, USA}

\author{Gregory F. Snyder}
\affiliation{Space Telescope Science Institute, 3700 San Martin Drive, Baltimore, MD 21218, USA}

\author[0000-0002-6748-6821]{Rachel S. Somerville}
\affiliation{Center for Computational Astrophysics, Flatiron Institute, 162 5th Avenue, New York, NY, 10010, USA}

\author[0000-0002-8770-809X]{Elizabeth R. Stanway}
\affiliation{Department of Physics, University of Warwick, Coventry, CV4 7AL, United Kingdom}

\author[0000-0002-4772-7878]{Amber N. Straughn}
\affiliation{Astrophysics Science Division, NASA Goddard Space Flight Center, 8800 Greenbelt Rd, Greenbelt, MD 20771, USA}

\author[0000-0002-8224-4505]{Sandro Tacchella}
\affiliation{Kavli Institute for Cosmology, University of Cambridge, Madingley Road, Cambridge, CB3 0HA, UK}\affiliation{Cavendish Laboratory, University of Cambridge, 19 JJ Thomson Avenue, Cambridge, CB3 0HE, UK}

\author[0000-0002-1410-0470]{Jonathan R. Trump}
\affiliation{Department of Physics, 196 Auditorium Road, Unit 3046, University of Connecticut, Storrs, CT 06269, USA}

\author[0000-0003-2338-5567]{Jes\'us Vega-Ferrero}
\affil{Instituto de Astrof\'isica de Canarias, La Laguna, Tenerife, Spain}

\author[0000-0003-3903-6935]{Stephen M.~Wilkins} 
\affiliation{Astronomy Centre, University of Sussex, Falmer, Brighton BN1 9QH, UK}
\affiliation{Institute of Space Sciences and Astronomy, University of Malta, Msida MSD 2080, Malta}

\author[0000-0001-8835-7722]{Guang Yang}
\affiliation{Kapteyn Astronomical Institute, University of Groningen, P.O. Box 800, 9700 AV Groningen, The Netherlands}
\affiliation{SRON Netherlands Institute for Space Research, Postbus 800, 9700 AV Groningen, The Netherlands}

\author[0000-0002-7051-1100]{Jorge A. Zavala}
\affiliation{National Astronomical Observatory of Japan, 2-21-1 Osawa, Mitaka, Tokyo 181-8588, Japan}

\begin{abstract}

We present a comprehensive analysis of the evolution of the morphological and structural properties of a large sample of galaxies at $z=3-9$ using early JWST CEERS NIRCam observations. Our sample consists of 850 galaxies at $z>3$ detected in both HST/WFC3 and CEERS JWST/NIRCam images, enabling a comparison of HST and JWST morphologies. We conducted a set of visual classifications, with each galaxy in the sample classified three times.  We also measure quantitative morphologies  across all NIRCam filters. We find that galaxies at $z>3$ have a wide diversity of morphologies. Galaxies with disks make up 60\% of galaxies at $z=3$ and this fraction drops to $\sim$30\% at $z=6-9$, while galaxies with spheroids make up $\sim30-40$\% across the redshift range and pure spheroids with no evidence for disks or irregular features make up $\sim20\%$. The fraction of galaxies with irregular features is roughly constant at all redshifts ($\sim40-50\%$), while those that are purely irregular increases from $\sim12\%$ to $\sim20\%$ at $z>4.5$. We note that these are apparent fractions as many observational effects impact the visibility of morphological features at high redshift. On average, Spheroid Only galaxies have a higher S\'ersic index, smaller size, and higher axis ratio than Disk or Irregular galaxies. Across all redshifts, smaller spheroid and disk galaxies tend to be rounder. Overall, these trends suggest that galaxies with established disks and spheroids exist across the full redshift range of this study and further work with large samples at higher redshift is needed to quantify when these features first formed.

\end{abstract}

\section{Introduction}\label{sec:intro}

Between the era of early galaxy formation and today, galaxies have undergone dramatic transformations in all respects. They have produced successive generations of stars from clouds of molecular gas, continuously building up their stellar populations, while enriching the interstellar medium with heavy elements. The gas reservoir within galaxies changed as they converted a fraction of their supply of cold molecular gas into stars and fresh gas was replenished via inflow from the intergalactic medium (IGM). The overall star formation rate (SFR) density of the universe grew until it reached a peak at $z \sim 2 - 3$ \citep{Madau14a} and then began to decline toward the present day low levels. The growth in the stellar mass of galaxies coincided with a change in their physical structure as the overall massive galaxy population transitioned from disk-dominated spiral galaxies into bulge-dominated elliptical galaxies. Throughout this assembly process, the central supermassive black holes (SMBHs) of galaxies grew, leading to an established relationship between SMBH and stellar mass (e.g., \citealt{McConnell12a}). 
Tracking the evolution of the structural properties of galaxies can provide key insights into the galaxy evolution pathways responsible for each of these transformations. Probing the different physical processes driving the formation of disks and bulges, the growth of SMBHs, the onset of star formation, and its subsequent cessation during a critical time in the universe's history is important for testing theoretical galaxy formation models. 

Deep extragalactic surveys with the {\it Hubble Space Telescope} (HST) have revolutionized our understanding of galaxy evolution between the peak epoch of galaxy assembly 10\,Gyr ago and today, but many open questions remain about the early phases of evolution within the first 3\,Gyr. When do we see the first disks in galaxies in the early universe? At what point did the first bulges form and do the physical processes responsible for their formation change with redshift? Does the quenching of star formation precede or follow the morphological transformation in these early galaxies? 

To robustly address these questions, it is essential to push our observations into the early universe, since most of our current understanding of galaxies at high redshift has come from galaxies at $z=1-3$, the period of time colloquially referred to as ``cosmic noon." Even though this represents a time period 10\,Gyr in the past, many galaxies at this time were already fairly mature and had structures, such as disks and bulges in star-forming galaxies, that generally resemble today's galaxies (e.g., \citealt{Tacchella15a,Costantin22a}). Previous large morphological studies of galaxies have typically been limited to galaxies at $z<3$ due to the fact that cosmological surface brightness dimming makes faint features in high redshift galaxies hard to detect and because the rest-frame optical emission that traces the broad stellar populations in galaxies is shifted beyond the wavelength capabilities of HST at higher redshifts. 

Early morphological studies with WFPC2 and ACS on HST were ground-breaking, quantifying for the first time the fraction of galaxies of various Hubble types (i.e., barred and unbarred spirals, ellipticals, and irregular galaxies) as a function of redshift, even beyond $z\sim1$ (e.g., \citealt{Abraham96a, Giavalisco1996, Lowenthal1997, con2000, Jogee2004, Elmegreen2004, Sheth2008, Lotz2006a,  Ravindranath2006}). However, at $z>1$, these optical surveys probed the rest-frame UV light of galaxies and found that very large fractions of distant galaxies had peculiar or clumpy morphologies, which suggested at the time that the Hubble sequence had not yet formed at these early times (e.g., \citealt{Abraham96a}). Investigations using near-infrared observations with NICMOS, sensitive to the rest-frame optical structure of galaxies, found that galaxies beyond $z\sim1$ presented a wide diversity of morphologies, including many objects that were compact or irregular, but also those that were morphologically mature spirals and ellipticals (e.g., \citealt{vandokkum01a,Stanford04a, Papovich05a, Daddi05a, Elmegreen2005}).

With the installation of WFC3 on HST in 2009, large samples of fainter galaxies at cosmic noon were observed. CANDELS, the Cosmic Assembly Near-infrared Deep Extragalactic Legacy Survey, \citep{Koekemoer11a,Grogin11a} obtained deep NIR imaging with WFC3 over a total of $\sim 800$ arcmin$^{2}$. These observations showed that while galaxies at $z\sim2$ were overall messier and clumpier, with larger fractions of mergers and generally irregular galaxies than today's universe, the general underpinnings of the Hubble sequence were already in place, i.e., a large fraction of star-forming galaxies were overall disk-like and passive galaxies are overall compact or spheroid-like (e.g., \citealt{Wuyts11a,Lee13a, Mortlock13a, vanderWel14a, Kartaltepe15a, Zhang19a}). This means that the first disks and spheroids must have begun to form at much earlier times.

With its unprecedented sensitivity in the infrared, the {\it James Webb Space Telescope} (JWST) is poised to make remarkable discoveries about this transformative era in galaxy assembly and test key theoretical predictions of our understanding of the physics of the early universe. The four pointings of deep multi-band NIRCam observations taken in 2022 June from the Cosmic Evolution Early Release Science (CEERS) survey (Finkelstein et al., in prep.) provide the first opportunity for a comprehensive analysis of the structural evolution of galaxies in the first 3\,Gyr of the Universe's history.

In this paper, we use these first CEERS observations to conduct an early analysis of the evolution of galaxy morphology and structure for a large sample of HST/WFC3-selected galaxies at $z=3-9$. This paper is organized as follows. In Section 2, we describe the basics of the CEERS observations and our data reduction, along with the ancillary data used to identify our sample of galaxies at $z>3$. In Section 3, we describe our morphological measurements, including visual classifications, parametric, and non-parametric morphologies. We present our results in Section 4 and discuss their implications in Section 5. Finally, we summarize our findings in Section 6. Throughout this paper, all magnitudes are expressed in the AB system and we assume a \cite{chabrier03} Initial Mass Function (IMF). We also assume the following cosmological parameters: $H_{0} = 70~{\rm km~s^{-1}~Mpc^{-1}}$, $\Omega_{tot} = 1$, $\Omega_{\Lambda} = 0.7$, and $\Omega_{M}=0.3$.

\section{Data}\label{sec:data}
\subsection{CEERS Observations and Data Reduction}\label{sec:reduction}

CEERS (Finkelstein et al., in prep.) is an Early Release Science (ERS) program (Proposal ID \#1345) that will observe the EGS (Extended Groth Strip, \citealt{davis07}) extragalactic deep field (one of the five CANDELS fields; \citealt{Koekemoer11a,Grogin11a}) early in Cycle 1 with data made available to the public immediately. CEERS will obtain observations in several different modes with JWST, including a mosaic of 10 pointings with NIRCam, NIRSpec multi-object spectroscopic observations in parallel for six pointings, and six pointings with MIRI in parallel.  The NIRCam imaging of CEERS will cover a total of 100 sq.\ arcmin with the F115W, F150W, F200W, F277W, F356W, F410M, and F444W filters down to a 5$\sigma$ depth ranging from 28.8--29.7 (for a typical total exposure time of 2835 s per filter). The first set of CEERS observations were taken on 2022 June 21 in four pointings, hereafter referred to as CEERS1, CEERS2, CEERS3, and CEERS6.

We performed an initial reduction of the NIRCam images in all four pointings using version 1.7.2 of the JWST Calibration Pipeline\footnote{\url{jwst-pipeline.readthedocs.io}} with some custom modifications. We used the current set of NIRCam reference files\footnote{\url{jwst-crds.stsci.edu}, jwst\_0989.pmap, jwst\_nircam\_0232.imap}, though we note that the majority were created pre-flight, including the flats. For details on the reduction steps, see \cite{Bagley2022a}. To summarize briefly, we applied detector-level corrections using Stage 1 of the pipeline with default parameters and used custom scripts to remove  $1/f$ noise, wisps, and snowballs from the count-rate maps. We then processed the cleaned count-rate maps with Stage 2 of the pipeline and then used a custom version of the TweakReg step to calibrate the astrometry. We then coadded the images using the drizzle algorithm with an inverse variance map weighting \citep{casertano00,fruchter02} in the Resample step in the pipeline. The final mosaics for each pointing in all filters have pixel scales of 0\farcs03/pixel. We then used a custom script to background subtract the images. These final background-subtracted images were used for the morphology measurements described in Section 3.

\subsection{CANDELS Images and Catalogs}\label{sec:candels}

For the analysis in this paper, before updated catalogs incorporating JWST photometry are available, we use existing CANDELS v2 redshifts and stellar masses for the HST F160W-selected galaxies in the EGS field. Here we provide a brief summary of these measurements; for full details on the photometric redshift measurements and resulting catalogs, see \cite{Kodra22a}. 

The v2 photometric redshifts and stellar masses are based on the CANDELS EGS photometric catalog of \cite{stefanon2017}, which includes broadband TFIT \citep{laidler07}  photometry from the UV to IR imaging from both ground- and space-based telescopes. For this paper, we adopt the \texttt{z\_best} column from the \cite{Kodra22a} catalog, which provides the overall best estimate of the redshift. This corresponds to the secure spectroscopic redshift if one is available or the \texttt{mFDa4\_z\_weight} photometric redshift otherwise. We use this photometric redshift value because it produces the most accurate confidence intervals (see \citealt{Kodra22a} for further details). 

We then use the \cite{stefanon2017} photometry and the above estimated redshift to determine stellar masses for galaxies in the CANDELS EGS field. This was done using two different codes: \texttt{Dense Basis}\footnote{\url{https://github.com/kartheikiyer/dense_basis}} and \texttt{P12}. \texttt{Dense Basis} \citep{iyer17, iyer19} is a python-based code that uses Flexible Stellar Population Synthesis (FSPS) to generate model spectra that correspond to a wide range of physically motivated non-parametric star formation histories (SFHs), metallicities, and dust attenuation values. \texttt{P12} \citep{Pacifici2012} is a Fortran-based code that uses a Bayesian fitting algorithm and model SEDs generated using simple stellar population models and the reprocessing of these models using the photo-ionization code \texttt{CLOUDY} \citep{Ferland1998}. For more details on each of these two methods and detailed comparisons, see \cite{iyer17}, \cite{iyer19}, \cite{Pacifici2012}, and Pacifici \etal (2022).

We find that the stellar masses using these two methods agree well with one-another, with the expected level of scatter (e.g., Pacifici \etal\ 2022). For this paper, we use the mean of the two measurements. We note here the caveats that these measurements are highly uncertain at the highest redshifts ($z>6$) and the faintest magnitudes (F160W$>26$) because they rely on the HST ACS and WFC3 photometry that do not trace the rest-frame optical light at these redshifts. Future work will improve on these measurements with the addition of JWST NIRCam fluxes.

In addition to the JWST CEERS images described above, we also use the existing CANDELS ACS and WFC3 images to compare the HST and JWST morphologies. Here we use the mosaics produced by the CEERS team\footnote{\url{https://ceers.github.io/releases.html\#hdr1}} with updated astrometry tied to Gaia-EDR3 and a pixel scale of 0\farcs03/pixel.

\subsection{Sample Selection}

For the sample analyzed in this paper, we select all galaxies with \texttt{z\_best}$>3$ within the CEERS1, CEERS2, CEERS3, and CEERS6 NIRCam SW or LW footprints. We cross-match these sources with those identified in NIRCam imaging using the \texttt{Source Extractor} segmentation map (see  \ref{sec:se} for more details). We remove any spurious sources in the NIRCam imaging, such as those that result from the diffraction spikes of stars and those that are so close to the edge of the images that we cannot obtain reliable morphology measurements. This results in a total sample of 850 sources with a detection in any of the NIRCam filters. We note here that at the magnitude limit of the CANDELS WFC3 images (F160W$<27.6$, 5$\sigma$) almost all of these galaxies have S$/$N in the NIRCam images high enough to enable morphological classifications, as discussed further in Section \ref{sec:measurements}. 

\begin{figure}
\includegraphics[scale=0.38]{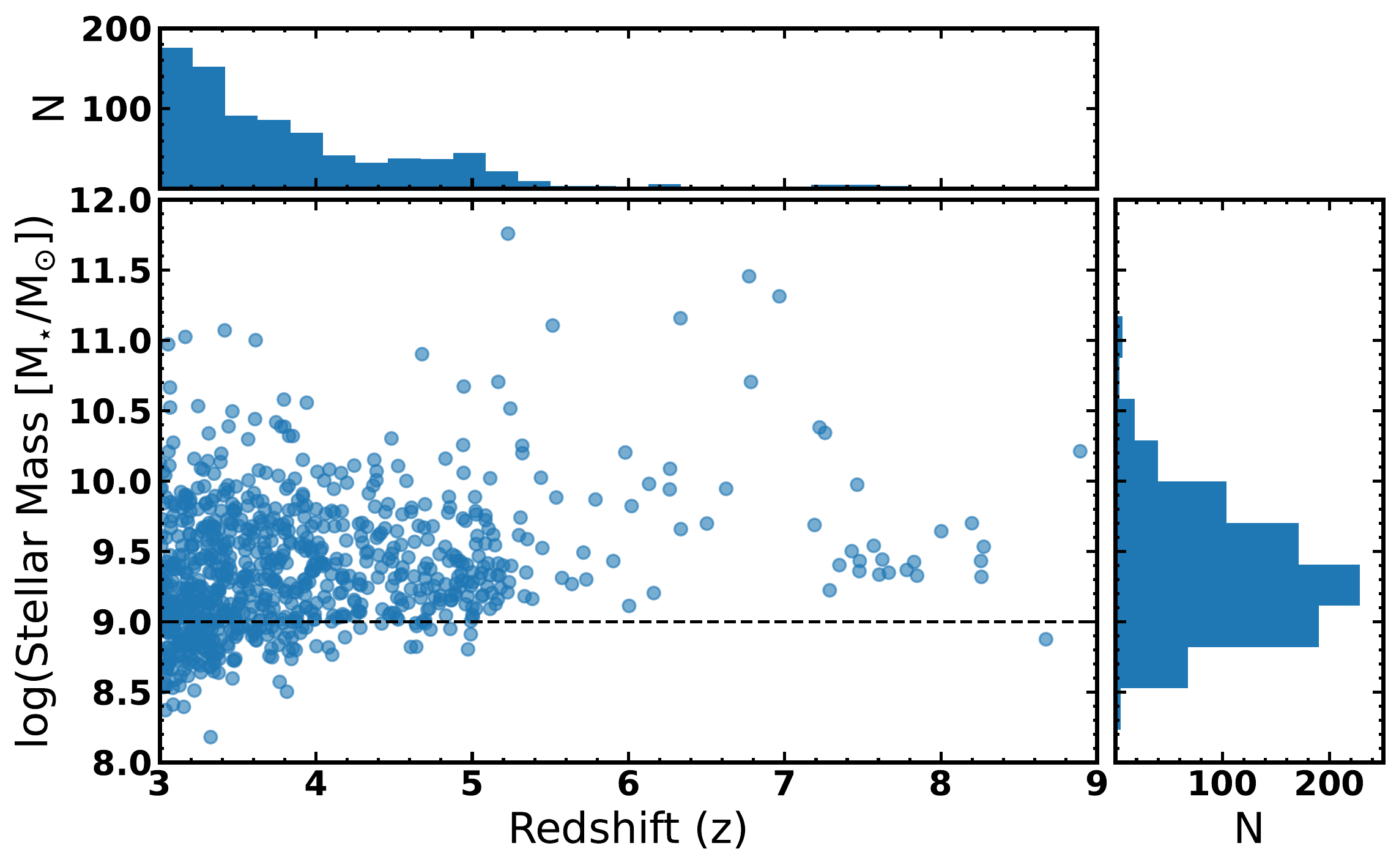}
\caption{Stellar mass versus redshift for the $z>3$ sample used in this paper. Above and to the right are the distributions of redshift and stellar mass, respectively. The horizontal dashed line at log(M$\star$/M$_{\odot}$)=9 corresponds to the mass cut used for the subsample of objects in Section \ref{sec:visual}.}
\label{zmass}
\end{figure}

Figure \ref{zmass} shows the redshift and mass distribution for this sample of objects. We note that ten of these objects have existing spectroscopic redshifts from either the MOSDEF \citep{kriek2015} or DEEP2 \citep{newman2013} spectroscopic surveys; we use these spectroscopic redshifts for these ten objects. Overall, this sample peaks at $z\sim3$ and has a long tail beyond $z\sim6$ out to $z\sim9$. There are a total of 40 sources in this sample with a photometric redshift of $z>6$. We note here that the redshifts and stellar masses for these sources are uncertain, and some of these may turn out to be at lower redshift. 
The redshift and stellar mass estimates will be improved with the addition of JWST data to the SED modeling in the future.

\section{Measurements}\label{sec:measurements}

\subsection{\texttt{Source Extractor} Setup}\label{sec:se}

Galaxies were detected in the NIRCam images using \texttt{Source Extractor}\footnote{\url{https://sextractor.readthedocs.io/}} version 2.25.0 \citep{ber1996}. The setup was optimized to detect the HST-selected $z>3$ galaxies without over-deblending. We created empirical PSFs for each filter by stacking stars and then the F115W, F150W, F200W, and F277W images were PSF-matched to the F356W image using the Python-based code \texttt{PyPHER}\footnote{\url{https://pypher.readthedocs.io/en/latest/}} \citep{Boucaud16a}. In order to detect galaxies that may be very faint in some of the NIRCam filters, we used an inverse-variance weighted combination of the PSF-matched F150W, F200W, F277W, and F356W images as the detection image. We first ran \texttt{Source Extractor} in a ``cold" mode to deblend nearby galaxies, then in ``hot" mode to detect faint galaxies, following \cite{stefanon2017}. We then combined the ``cold" and ``hot" detections in order to keep all objects that were detected by at least one mode, using a factor of 2.5 to enlarge the cold isophotes. We visually inspected the segmentation map for the $z>3$ sources and optimized the parameters to ensure that the sources were detected and adequately deblended from nearby neighbors without being shredded. We use the final segmentation map produced from this process for all of the measurements presented in this section.

\subsection{Visual Classifications}

Each of the 850 galaxies in our $z>3$ sample were classified by three different people from among a total of 35 members of our team. We used the Zooniverse project builder\footnote{\url{https://www.zooniverse.org/lab}} to host images of each galaxy and designed a workflow of five tasks based on a modified version of the classification scheme of \cite{Kartaltepe15a}. These five tasks ask classifiers to select options for the galaxy's main morphology class, their interaction class, various structural and quality flags, and finally to leave any specific comments about a particular object. In this paper, we focus on the first of these five tasks, the main morphology classification, which roughly corresponds to the typical Hubble type classification. The options for the main morphology class are: 1) Disk, 2) Spheroid, 3), Irregular / Peculiar, 4) Point Source / Unresolved, and 5) Unclassifiable / Junk. To reflect the overall diversity seen in high redshift galaxies, these classes are not mutually exclusive, so a classifier can choose multiple options to best reflect the overall morphology of the galaxy. For example, a galaxy can have both a disk and a spheroid in the case that it is a disk galaxy with a bulge component. A galaxy can be both a disk and irregular, if for example it is an asymmetric disk or a disk involved in an interaction. The exception is that if a galaxy is `Unclassifiable' then it cannot also be one of the other classes. This level of complexity can make the interpretation of the various classes challenging, but the extra information provides an important level of nuance to the classifications.

Classifiers are presented with a collection of postage stamps for each galaxy being classified. These stamps include each of the JWST NIRCam filters (F115W, F150W, F200W, F277W, F356W, F410M, and F444W) at their native resolution, with an asinh scaling to bring out low surface brightness features an RGB stamp made up of the filters that correspond to the rest-frame optical, a version of that stamp zoomed out by a factor of two, the NIRCam \texttt{Source Extractor} segmentation map described above, three HST ACS/WFC3 filters (F814W, F125W, and F160W, also with an asinh scaling), and finally an RGB stamp of these three HST filters and a similarly zoomed out version. The stamps are scaled by the size of the galaxy as measured by \texttt{Source Extractor}, following Equations 2 and 3 of \cite{Haussler2007}, with a minimum size of 100$\times$100 pixels. An example set of stamps for one of the galaxies is shown in the Appendix.

The classifiers are asked to make a holistic judgement about the overall morphology of the galaxy, taking information across the full wavelength range into account. In a separate task, the classifiers can select flags to indicate that the morphology changes across the NIRCam filters or differs between JWST and HST images.

\subsection{Parametric Fits}\label{sec:fits}

We perform parametric fits on the NIRCam images using both \texttt{Galfit} \footnote{\href{https://users.obs.carnegiescience.edu/peng/work/galfit/galfit.html}{https://users.obs.carnegiescience.edu/peng/work/galfit/galfit.html}}  \citep{peng2002,peng2010} and \texttt{GalfitM} \footnote{\href{https://www.nottingham.ac.uk/astronomy/megamorph/}{https://www.nottingham.ac.uk/astronomy/megamorph/}}  \citep{hau2013}. 
\texttt{Galfit} is a least-squares fitting algorithm that finds the optimum S\'ersic fit to a galaxy's light profile and \texttt{GalfitM} is a modified version that uses images at different wavelengths to allow one to constrain the fit parameters that vary smoothly as a function of wavelength. The benefit of using \texttt{GalfitM} is that it fits all bands simultaneously and produces more physically consistent models. We performed fits using both codes to test for self-consistency, but since we use the rest-frame optical fit throughout this paper, we focus here on the \texttt{GalfitM} fits and describe the \texttt{Galfit} fits in the Appendix.

\texttt{GalfitM} fits were performed using the IDL program \texttt{Galapagos-2} from the \texttt{MegaMorph} Project\footnote{\href{https://www.nottingham.ac.uk/astronomy/megamorph/}{https://www.nottingham.ac.uk/astronomy/megamorph/}}\citep{bam2011,vika2013,hau2013,Haussler2022}. \texttt{Galapagos-2} is a wrapper that enables \texttt{GalfitM} to be run over larger survey images. We used the \texttt{Source Extractor} setup described above with \texttt{Galapagos-2}. As input, we used all seven NIRCam filters (F115W, F150W, F200W, F277W, F356W, F410M, and F444W) and used the NIRCam WHT images produced by the JWST pipeline to create RMS images to be used as the input sigma image. We used the F200W \texttt{Source Extractor} catalog for initial guesses and used the NIRCam empirical PSFs. In addition to the final output catalog with the S\'ersic fit parameters, \texttt{Galapagos-2} also outputs the original stamp, the \texttt{GalfitM} model, and the residual image for each galaxy in each filter. Out of the 850 $z>3$ galaxies in our sample, 37 (4\%) were flagged because \texttt{GalfitM} reached one of the constraint limits in one of the filters.

\subsection{Non-parametric Measurements}\label{sec:statmorph-fits}

We measure non-parametric morphologies using the Python package \texttt{Statmorph}\footnote{\url{https://statmorph.readthedocs.io/en/latest/}} \citep{Rodriguez-Gomez19a}. For each NIRCam filter, we create 100$\times$100 pixel cutouts of the 850 galaxies in our $z>3$ sample to use as input to \texttt{Statmorph}, along with a cutout of the segmentation map, the empirical PSF, and the gain. 

\texttt{Statmorph} measures a wide range of morphology statistics commonly used in astrophysics. The ones that we use for the analysis in this paper are: concentration ($C$), asymmetry ($A$), and clumpiness/smoothness ($S$) \citep{ber2000,con2000,con2003}; the Gini coefficient ($G$) and the second moment of the region of the galaxy containing 20\% of the total flux ($M_{20}$) \citep{abra2003,lotz2004}; the Gini-M$_{20}$ bulge, and merger statistics \citep{Rodriguez-Gomez19a}, 
the signal-to-noise per pixel, and quality flags. 

As for the parametric fits, we use the fits for the filter corresponding to the rest-frame optical emission at the redshift of the galaxy: F277W for galaxies at $z=3.0-4.0$, F356W for galaxies at $z=4.0-4.5$, and F444W for galaxies at $z>4.5$. Of the 850 galaxies fit, 81\% have a signal-to-noise per pixel of $> 2.5$ in the corresponding rest-frame optical filter; below this value, the fit results may not be reliable \citep{Lotz2006a}. We compare these commonly used measures of galaxy morphology to our visual classifications in Section \ref{sec:statmorph}.

\begin{figure}
\hspace*{-0.1in}
\includegraphics[scale=0.65]{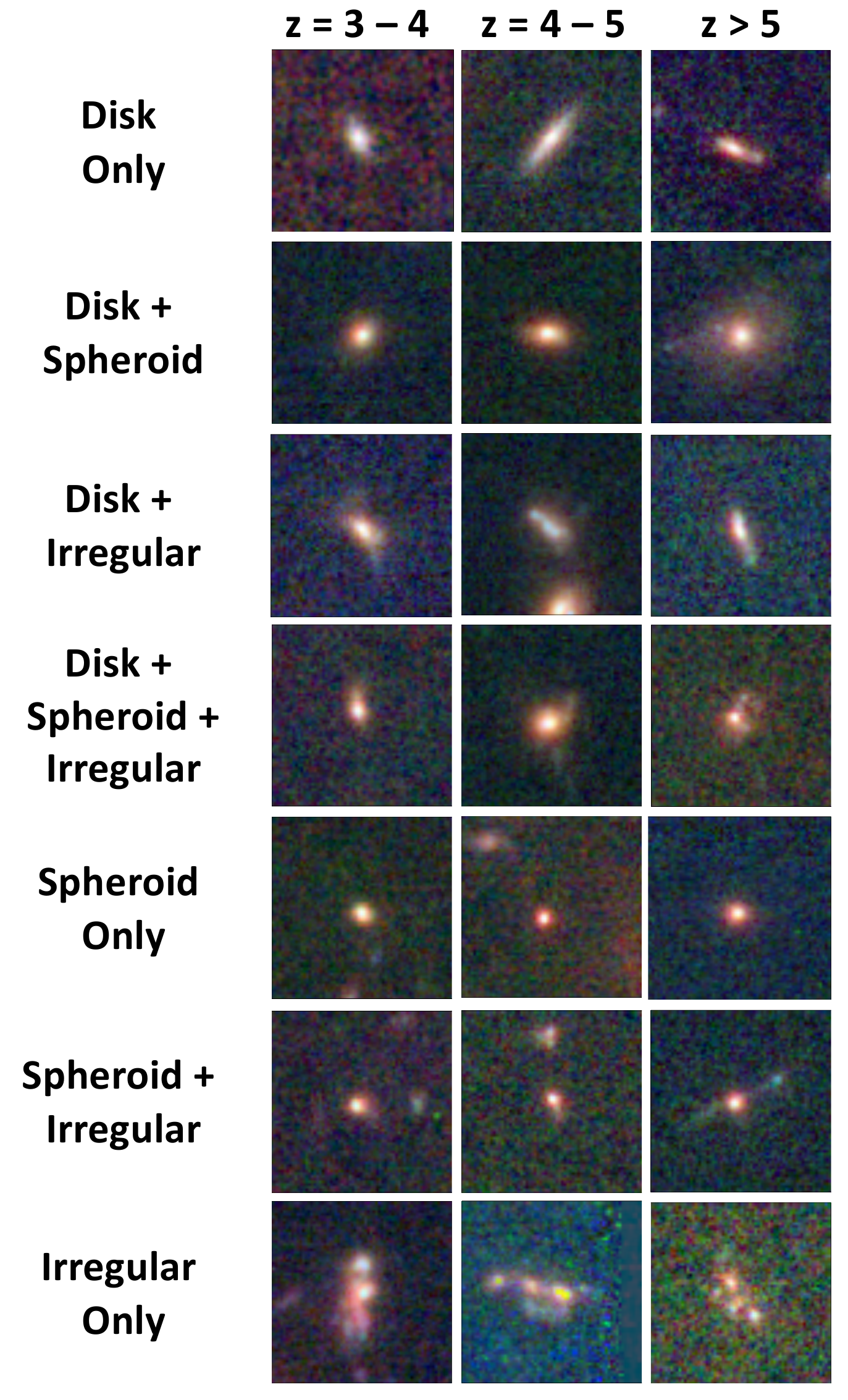}
\caption{NIRCam F150W+F277W+F356W postage stamp cutouts of a selection of example galaxies in each of the seven morphological groups described in Section\ref{sec:visual} at three different redshift bins. Each cutout is 2$''$ on a side.}
\label{fig:examples}
\end{figure}

\section{Results}\label{sec:results}

\subsection{Visual Classifications}\label{sec:visual}

For each of the 850 galaxies in our $z>3$ sample, we assign a galaxy a given visual classification if two out of three people select a given option as the main morphological class.  There is only one object in our sample for which all three classifiers disagree, meaning one selected only `disk', one selected only `spheroid', and one selected only `irregular. This object is therefore not included in any of the figures presented here. As noted above, since the main morphological classes are not mutually exclusive, various combinations are possible. Throughout this paper, we break things down into the following non-exclusive morphological groups (highlighted in Figure \ref{fig:examples}):

\begin{figure*}
\includegraphics[scale=0.78]{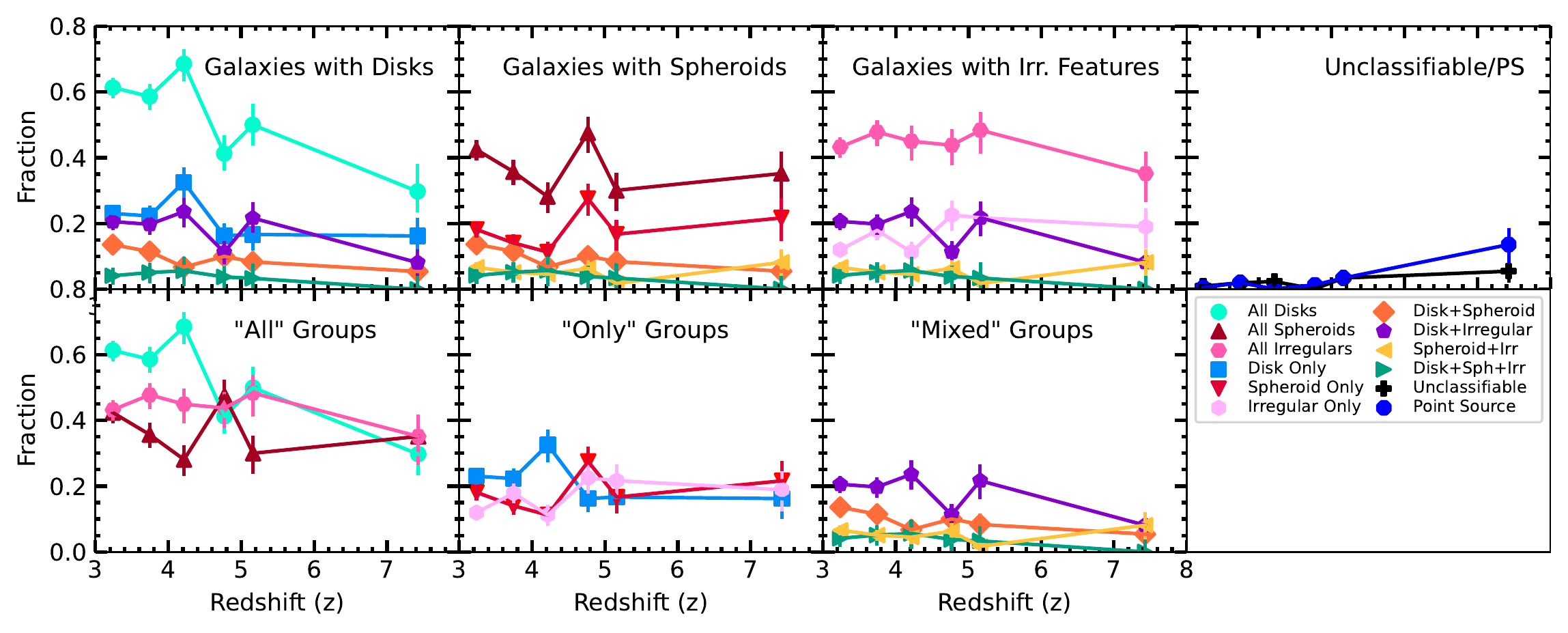}
\caption{The fraction of $z>3$  galaxies detected by both JWST and HST with M$_{\star}>10^{9}$\,M$_{\odot}$ as a function of redshift for each morphology class. The top row, from left to right shows galaxies with disks, galaxies with spheroids, galaxies with irregular features, and Point Sources and Unclassifiable galaxies. The bottom row shows all of the same morphological groups, but divided in different ways for easy comparison. From left to right, the combination of all disks, all spheroids, and all irregulars; the combination of Disk Only, Spheroid Only, and Irregular Only groups; and finally, the remaining mixed groups. Error bars represent the 1$\sigma$ binomial confidence limits given the number of objects in each category, following the method of \cite{cam2011}.}
\label{fig:fractions_z}
\end{figure*}

\begin{itemize}
    \item {\it Galaxies with Disks:} The Disk category contains galaxies classified as Disk Only (without a spheroid or irregular component), Disk+Spheroid (a galaxy with both a disk and spheroid component; a separate structural flag indicates whether the disk or the bulge is dominant), Disk+Irregular (a disk galaxy with irregularities such as asymmetries, a warp, or disturbance by a nearby companion), Disk+Spheroid+Irregular (a disk galaxy with a spheroid component that also has some irregularities; note that these are fairly rare). When we refer to `All Disks' we are referring to the sum of the galaxies in all of these categories.
    
    \item {\it Galaxies with Spheroids:} The Spheroid category contains galaxies classified as Spheroid Only (without a disk or irregular component), Spheroid+Disk (same as Disk+Spheroid above), Spheroid+Irregular (a spheroid galaxy with irregularities such as asymmetries or surrounding low surface brightness features), Spheroid+Disk+Irregular (same as Disk+Spheroid+Irregular above). When we refer to `All Spheroids' we are referring to the sum of the galaxies in all of these categories.

    \item {\it Galaxies with Irregular Features:} The Irregular category contains galaxies classified as Irregular Only (no discernible disk or spheroid component), Irregular+Disk (same as Disk+Irregular above), Irregular+Spheroid (same as Spheroid+Irregular above), Irregular+Disk+Spheroid (same as Disk+Spheroid+Irregular above). When we refer to `All Irregulars' we are referring to the sum of the galaxies in all of these categories. Note that the Irregular category may include merging or interacting systems, but also galaxies that are irregular for other reasons, such as clumpy star formation. Mergers and Interactions themselves will be discussed in a future paper (Rose et al., in prep).

\end{itemize}

Our sample contains the full range of morphological types across all redshift and stellar masses. Over the entire redshift range, only 16 and 18 galaxies are classified as Point Source/Unresolved or Unclassifiable, respectively. Figure \ref{fig:fractions_z} shows the fraction of the total number of galaxies that each morphological class makes up as a function of redshift. For a fair comparison across redshifts, we limit this to the 666 galaxies with stellar masses greater than $10^9$\,M${_\odot}$ since the galaxies with lower stellar masses are only present at the low redshift end  of our sample (see Fig.~\ref{zmass}). We emphasize that this represents a mass-selected sample but not a mass-complete sample, as there are likely to be many galaxies identified by JWST in this mass range that are undetected by CANDELS HST imaging.

Overall, 56\% of the galaxies above this mass cut at $z>3$ have a visually identifiable disk component, dropping from $\sim 60\%$ at $z=3-4$, to $\sim 45\%$ at $z\sim5$, to $\sim30\%$ at $z>6$. Again, we note that the photometric redshifts at $z>6$ are uncertain, so some fraction of these sources may actually be at lower redshift; we caution the reader when interpreting the results at $z>6$. The Disk Only and the Disk+Irregular groups each make up $\sim20\%$ (and slightly less at $z>6$) while the Disk+Spheroid group makes up $\sim10\%$ and the Disk+Spheroid+Irregular group makes up $<5\%$. 38\% of the galaxies at $z>3$ have a visually identifiable spheroidal component, decreasing from 42\% to 26\% between $z=3$ and 4.5, then varying between $\sim$30-40\% beyond $z=4.5$. This is largely driven by the similar decrease then increase in the Spheroid Only group. Part of this apparent trend at higher redshifts may be due to small number statistics and part may be due to a number of selection effects. For example, there is a possibility that we miss fainter extended features in some of these systems at high redshift. It is also possible that a larger fraction of galaxies at higher redshift are small enough to be at the resolution limit of NIRCam, given the expected size evolution of galaxies, and are therefore more round and compact in appearance. 

\begin{figure*}
\includegraphics[scale=0.55]{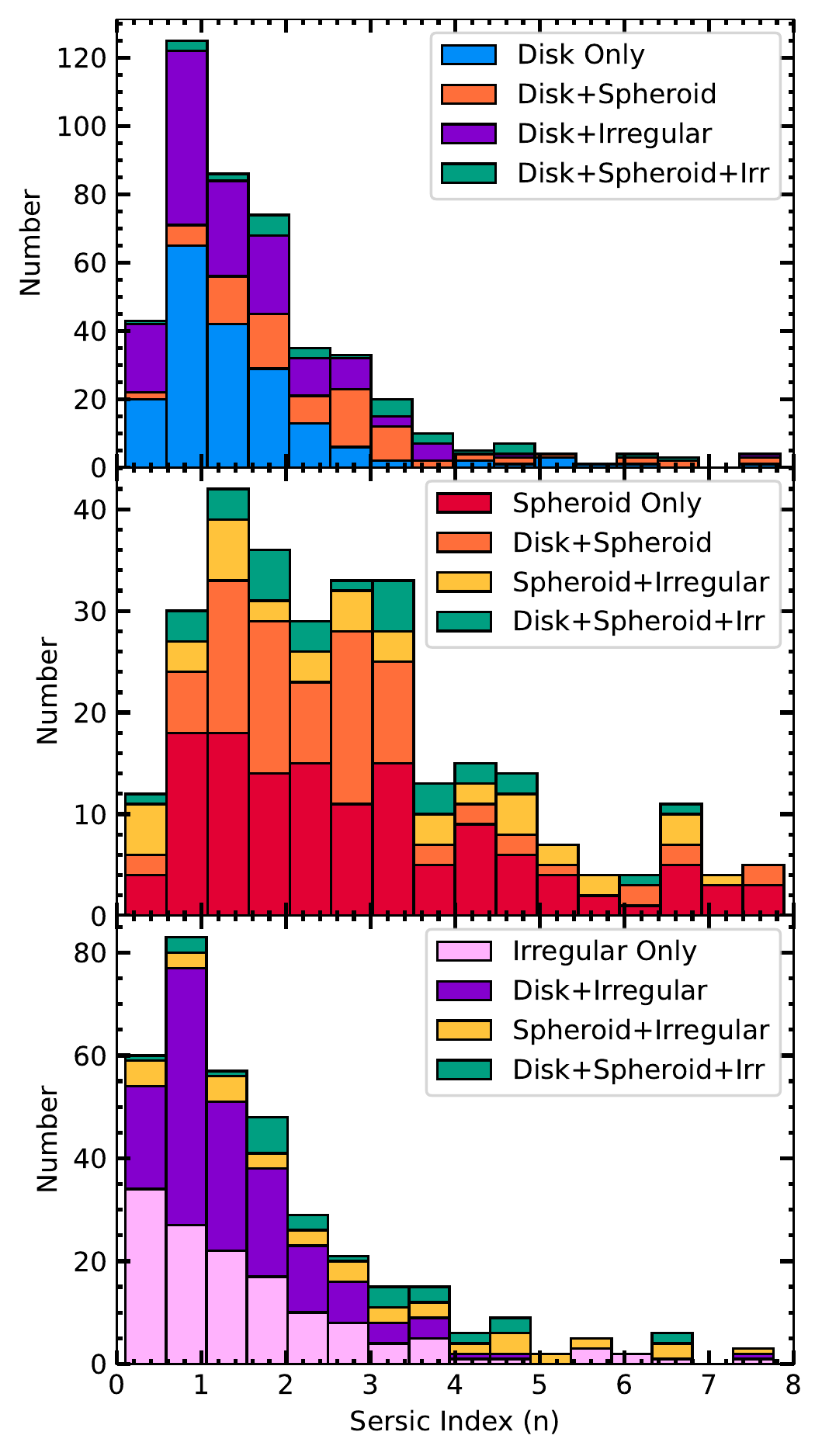}
\includegraphics[scale=0.55]{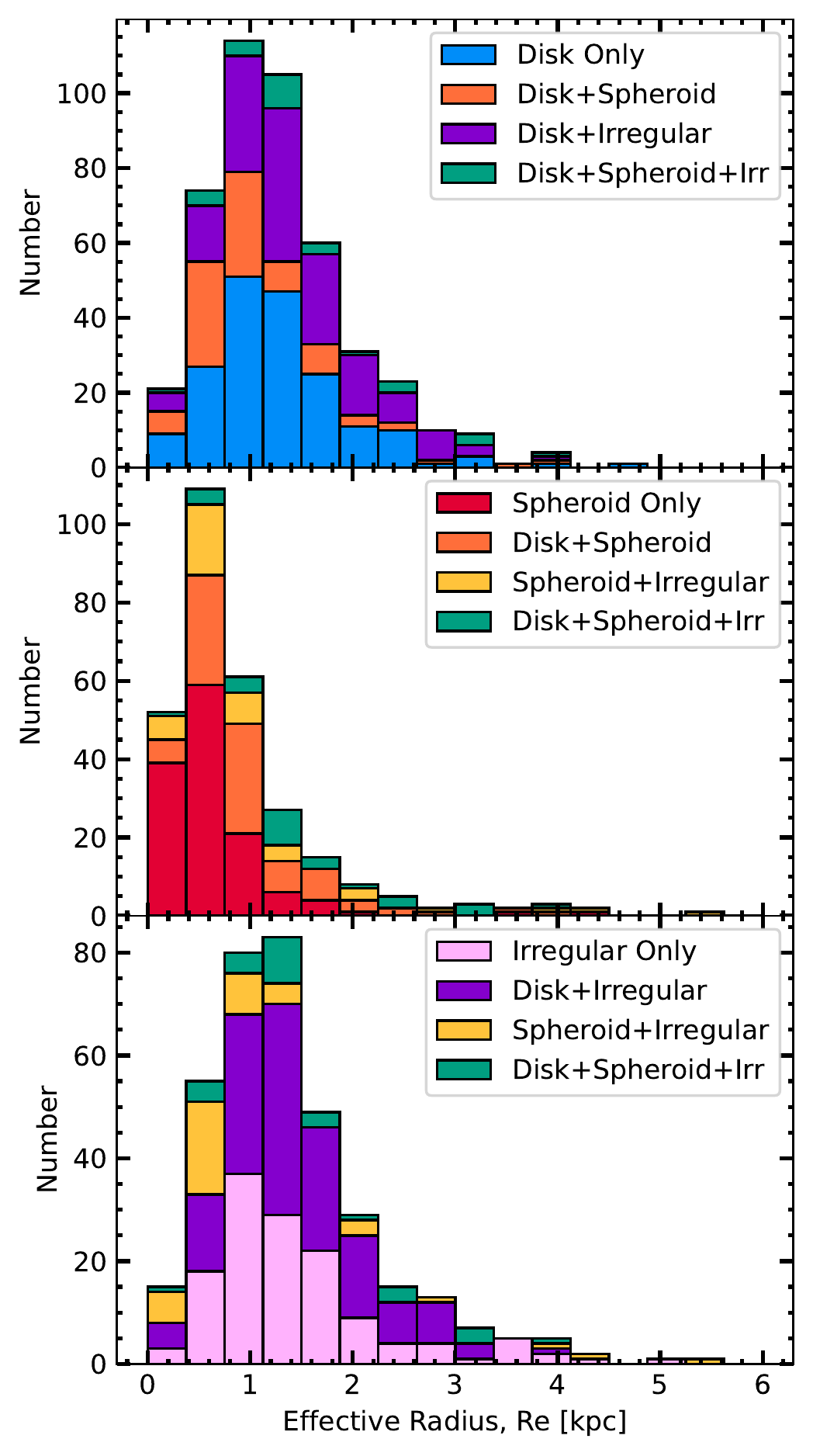}
\includegraphics[scale=0.55]{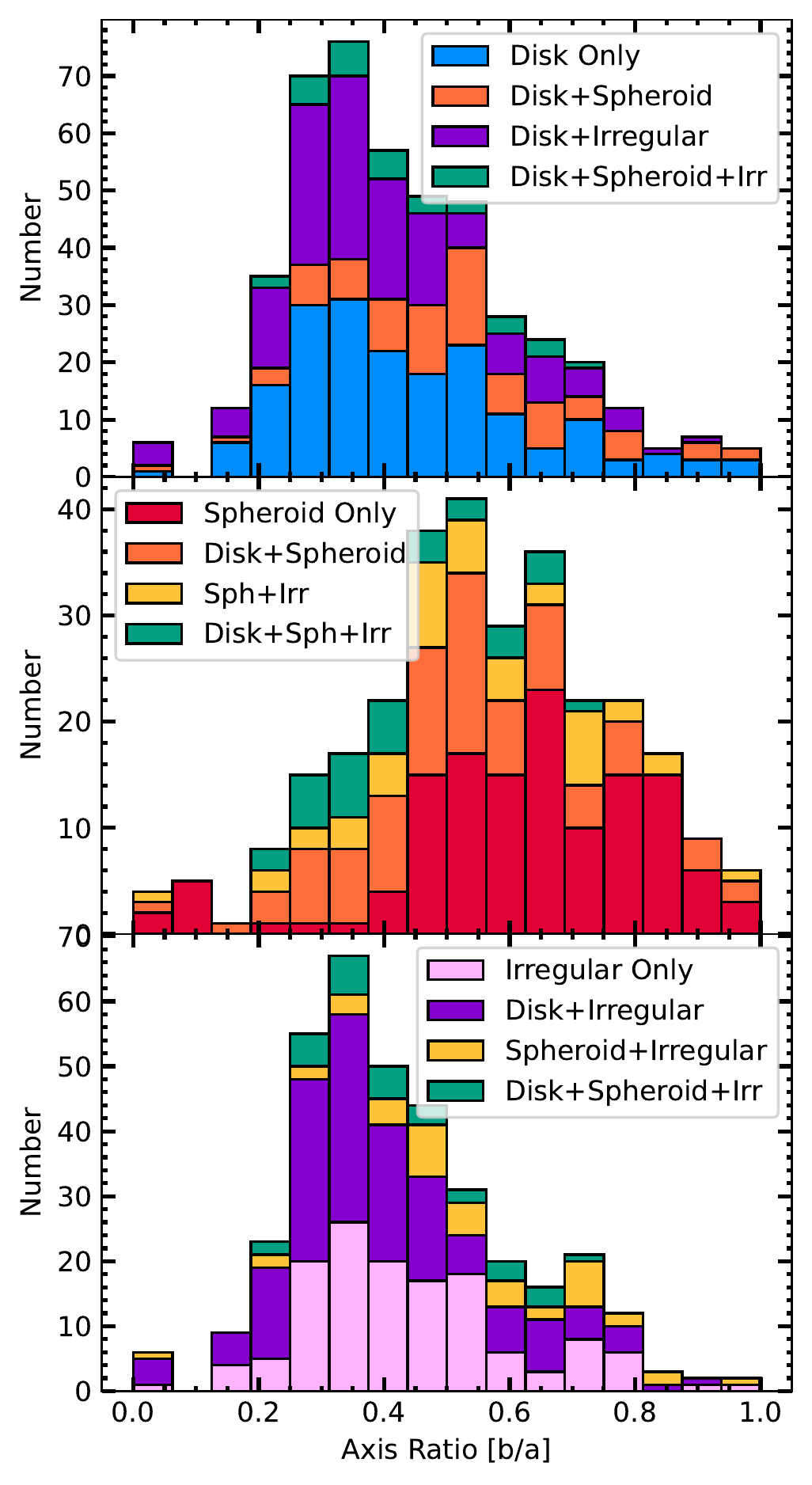}
\caption{Stacked histograms illustrating the distributions of the rest-frame optical S\'ersic Index ($n$), effective radius (R$_{e}$), and axis ratio ($b/a$) for the $z>3$ galaxy sample. The colors indicate the different combinations of the main morphological class chosen by two out of three people during the visual classifications, as described in Section \ref{sec:visual} and Figure \ref{fig:examples}.}
\label{fig:galfit}
\end{figure*}

43\% of the galaxies at $z>3$ have irregular features and this fraction remains roughly constant across the full redshift range due in part to the fraction of Disk+Irregular galaxies being roughly constant at $20\%$ and then decreasing while the fraction of Irregular Only galaxies is at roughly $10-15$\% and then increases to $20\%$ by $z=4.5$. Note that the total fractions of objects that are All Disks, All Spheroids, or All Irregular do not add up to one due to the overlapping objects in each of these classes.

Finally, we note that the fraction of point sources and unclassifiable objects remains at $0-2\%$ across most of the redshift range. At $z>6$, 13\% of galaxies are unresolved and 8\% are unclassifiable, corresponding to 5 and 3 individual galaxies, respectively, in this redshift bin. We remind the reader that the above percentages correspond to galaxies that were bright enough to be detectable with HST CANDELS imaging and may not be representative of the overall galaxy sample detectable by JWST at these redshifts.

\subsection{Comparison with Surface Brightness Profile Fits }\label{sec:galfit}

\begin{figure*}
\centering
\includegraphics[scale=0.85]{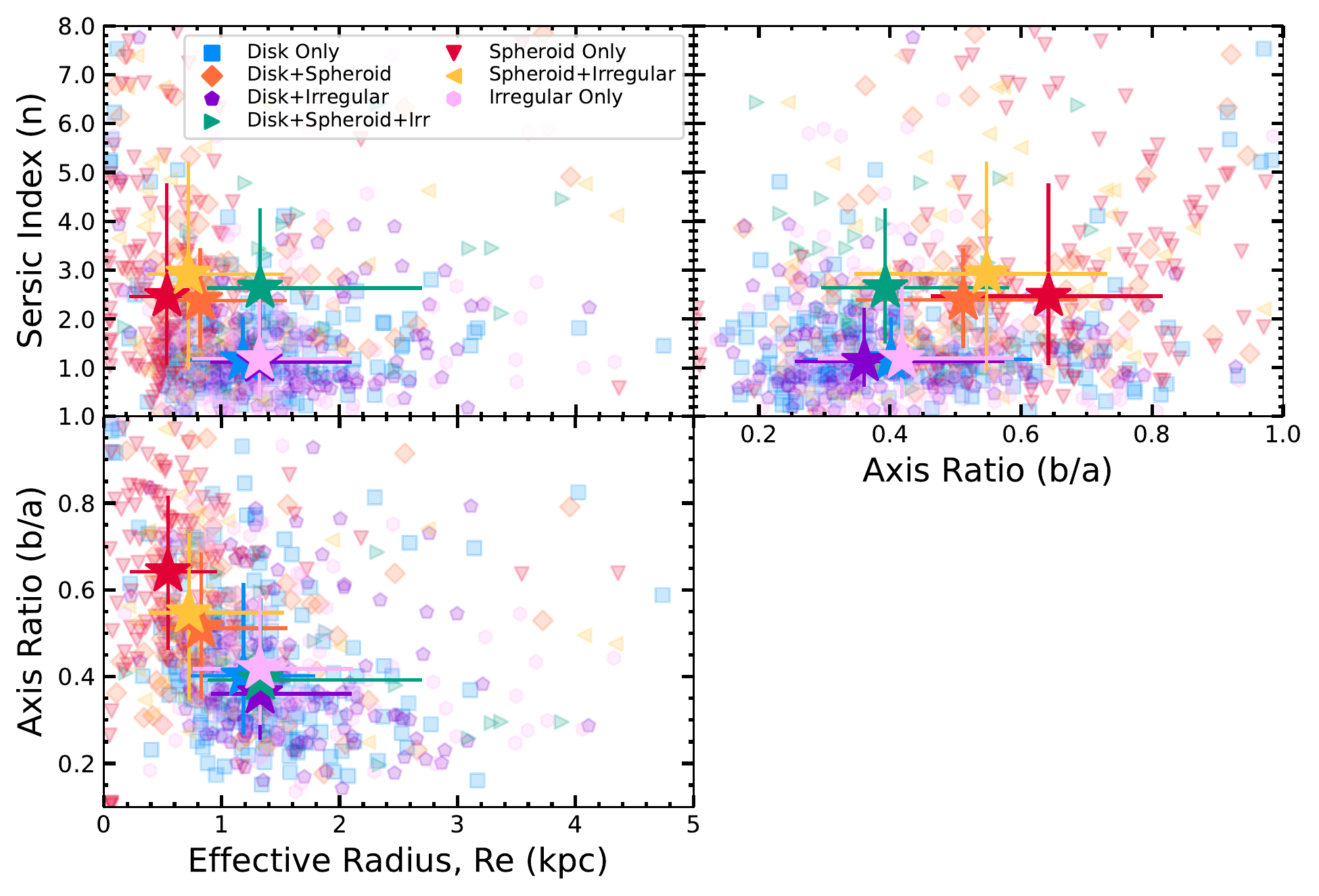}
\caption{The rest-frame optical S\'ersic index ($n$) plotted as a function of the effective radius (R$_{e}$) in kpc (top left) and the axis ratio ($b/a$) (top right) for the $z>3$ galaxy sample. The axis ratio plotted as a function of the effective radius is shown on the bottom left. The colors of each point indicate the different combinations of the main morphological class chosen by two out of three people during the visual classifications, as described in Section \ref{sec:visual} and Figure \ref{fig:examples}. The median values for each group are shown as stars with the error bars representing the 16th-84th percentile range of the distribution.}
\label{fig:n_Re_Q}
\end{figure*}

One of the major advances that JWST NIRCam imaging brings to morphological analyses of galaxies is that the broad wavelength coverage enables us to probe the rest-frame optical morphologies of galaxies across a wide redshift range. As described in Section \ref{sec:fits}, we used \texttt{GalfitM} to perform multiwavelength parametric fits across all of the NIRCam filters. For a fair comparison of these parameters at different wavelengths, we select the NIRCam filter closest to the rest-frame optical at different redshifts. Throughout this section, we use the F277W filter for galaxies at $3.0<z<4.0$, F356W for galaxies at $4.0<z<4.5$, and F444W for galaxies at $z>4.5$. 
We compare these measurements to the visual classifications in Figures \ref{fig:galfit} and \ref{fig:n_Re_Q}. 

Overall, the distribution of S\'ersic indices (where $n = 0.5$ corresponds to a Gaussian profile, $n = 1$ to an exponential profile, and $n = 4$ to a de Vaucouleurs profile)  tracks with the expectations from the visual classifications. Disk galaxies with no apparent spheroid or irregular features (Disk Only) peak at low S\'ersic indices with a long tail out to higher values, as expected ($\langle n\rangle=1.16^{+0.88}_{-0.40}$, where the error bar denotes the 16th-84th percentile range of the distribution). Galaxies that are pure spheroids (Spheroid Only) have a much broader distribution and peak at higher $n$ ($\langle n\rangle=2.46^{+2.32}_{-1.41}$) as has been noted at lower redshift based on HST imaging (e.g., \citealt{vika2015,Kartaltepe15a}). Galaxies with both a disk and spheroidal component (Disk+Spheroids) peak at intermediate values ($\langle n\rangle=2.39^{+1.06}_{-0.98}$). This illustrates that a cut at a fixed S\'ersic index would not cleanly select disk or spheroidal dominated galaxies. For example, a dividing line of $n=2$ would identify 71\% of the visually identified disks and only 45\% of the visually identified spheroids. However, it is worth noting that that a fraction of the objects visually identified as spheroids with low $n$ might have extended low surface brightness disks that are difficult to pick out by eye.

Irregular galaxies with no disk or spheroid component peak at very low $n$, with a substantial fraction at $n<1$ and a long tail out to higher values ($\langle n\rangle= 1.19^{+1.51}_{-0.82}$). Disk galaxies with irregular features (Disk+Irregular) peak closer to $n=1$ with a narrower distribution that more closely resembles that of the Disk Only galaxies. Likewise, the distribution of S\'ersic indices for the spheroidal galaxies with irregular features (Spheroid+Irregular) closely resembles that of the Spheroid Only group. A visual inspection of the models and residuals for the irregular galaxy population reveals that, unsurprisingly, irregular features are not well-fit by a S\'ersic profile. For disks and spheroidal galaxies with irregular features, the model fits the disk/spheroidal component well and leaves behind features in the residuals, while the irregular only population are not well-fit at all. We caution the reader against over-interpreting S\'ersic indices for irregular galaxies and against using S\'ersic indices to select disk galaxies without first checking the images (and residuals) for irregular features.

\begin{figure*}
\centering
\includegraphics[scale=0.85]{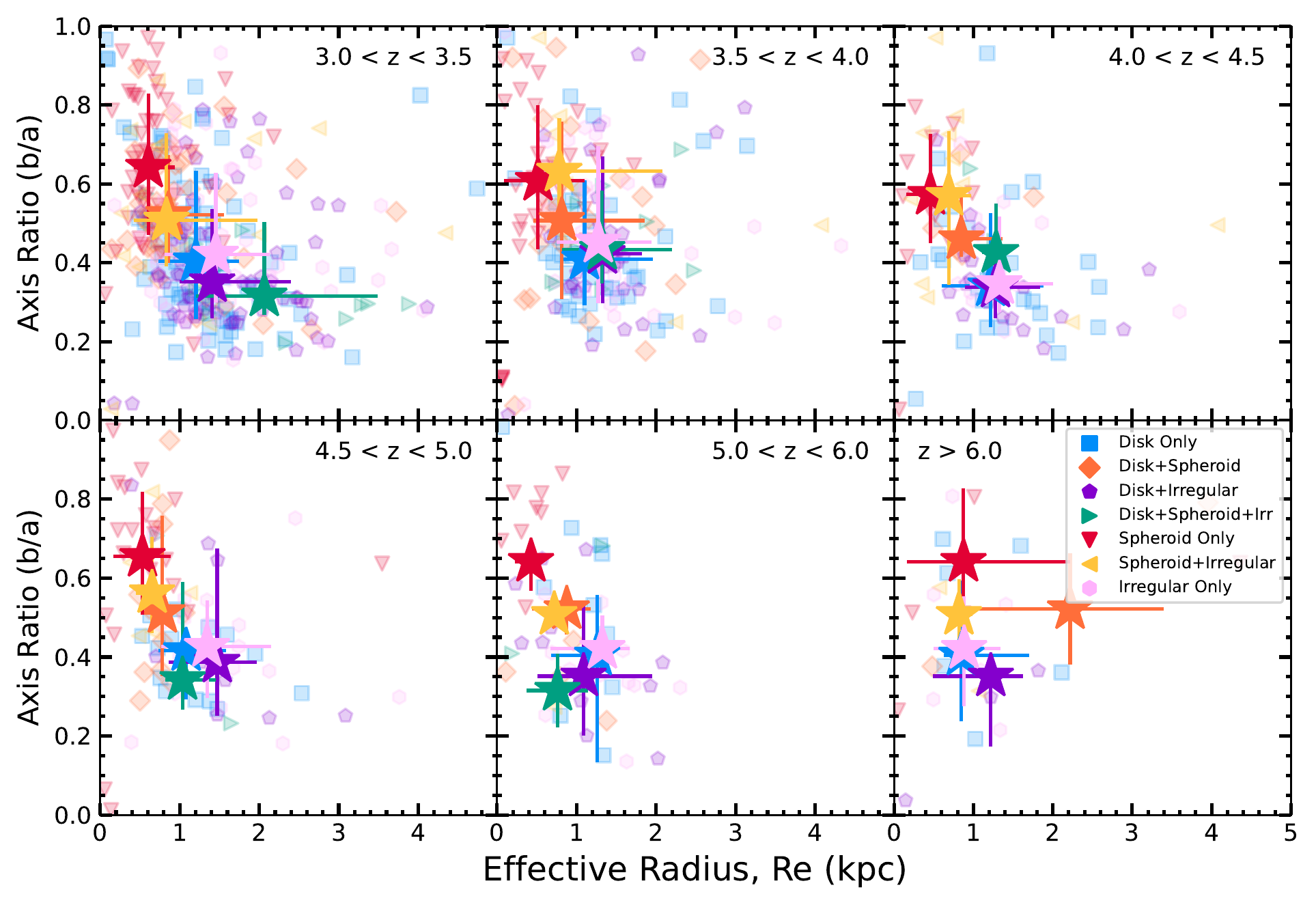}
\caption{The rest-frame optical axis ratio ($b/a$)  plotted as a function of the effective radius (R$_{e}$) in kpc in six different redshift bins.  The colors indicate the different combinations of the main morphological class chosen by two out of three people during the visual classifications, as described in Section \ref{sec:visual} and Figure \ref{fig:examples}. The individual sources are shown as transparent circles and the median for each group is shown as a star with the 16th-84th percentile range of the distribution shown as error bars.}
\label{fig:Q_Re_Redshift}
\end{figure*}

The center panel of Figure~\ref{fig:galfit} shows the distribution of sizes (effective radii, R$_{e}$) measured by \texttt{GalfitM} for each of the morphological types. Galaxies with disks and irregular features generally have larger sizes than those with spheroids. For example, galaxies with disks only have a median effective radius of 1.19$^{+0.61}_{-0.48}$\,kpc, irregular only have $\langle$R$_{e}\rangle=1.32^{+0.79}_{-0.57}$\,kpc, while spheroid only galaxies have $\langle$R$_{e}\rangle=0.54^{+0.42}_{-0.32}$\,kpc. These trends are seen more clearly in Figure \ref{fig:n_Re_Q}. The Disk+Irregular and Disk+Spheroid+Irregular groups have size distributions that more closely match the distribution for the Irregular Only group, while the Spheroid+Irregular group has a smaller median size ($\langle$R$_{e}\rangle=0.74^{+0.81}_{-0.34}$\,kpc). 

The distribution of the axis ratios is shown in the right panel of Figure \ref{fig:galfit} and in \ref{fig:n_Re_Q} and offers another way to compare our visual morphologies to a quantitative measurement. A population of disks with exponential profiles and random orientations is expected to have a relatively flat distribution of axis ratios that falls off at low values, while triaxial ellipsoids are expected to have a distribution that is peaked at higher values, $b/a \sim 0.6$, (e.g., \citealt{Elmegreen2005, Ravindranath2006, Padilla2008, Law2012, Robertson2022b}). The mean values we see for the different morphological groups follows this trend. The Spheroid Only group has the largest median axis ratio ($0.64^{+0.17}_{-0.18}$), while the Disk+Irregular group has the smallest (0.36$^{+0.21}_{-0.11}$).

Figure \ref{fig:Q_Re_Redshift} shows the axis ratio as a function of effective radius split into several redshift bins. In each redshift bin from $z=3$ to $z=6$, the Spheroid Only galaxies have the smallest median effective radius and the largest median axis ratios, suggestive of a population of true triaxial ellipsoids. Overall, we see the general trend of small galaxies being rounder, for all morphology types, as seen by \cite{Padilla2008} and \cite{Zhang19a}, in each of these redshift bins. We cannot draw any conclusions at $z>6$ due to the small sample size and the previously mentioned uncertainties.

\begin{figure*}
\hspace*{-0.2in}
\includegraphics[scale=0.7]{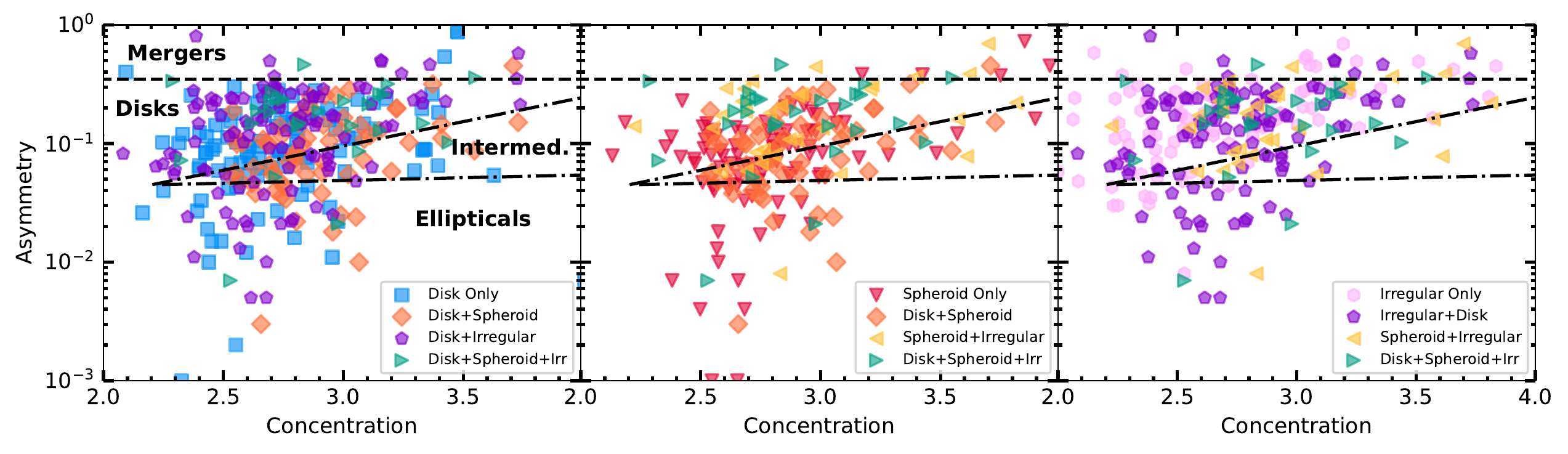}
\includegraphics[scale=0.7]{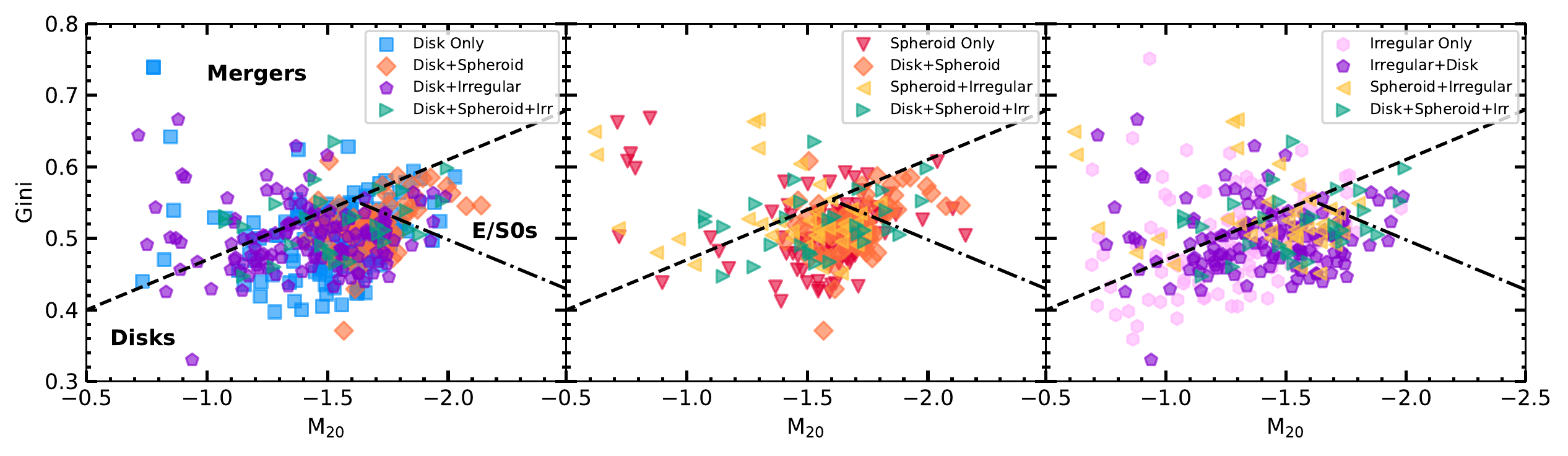}
\caption{{\it Top}: The rest-frame optical asymmetry value as a function of concentration for all galaxies at $z>3$ split by all disk galaxies (left), spheroids (center), and irregulars (right).  The dash-dotted lines show the boundaries between disk galaxies, elliptical galaxies, and intermediate galaxies from \cite{ber2000} and the dashed line is the dividing line above which nearby galaxies are expected to be major mergers ($A=0.35$; \citealt{con2003}). {\it Bottom}: The rest-frame optical Gini value as a function of M$_{20}$ for all galaxies at $z>3$ split by all disk galaxies (left), spheroids (center), and irregulars (right). The lines show the boundaries between disk and elliptical galaxies (dash-dotted), and mergers (dashed) from \cite{lotz2008}. The colors indicate the different combinations of the main morphological class chosen by two out of three people during the visual classifications, as described in Section \ref{sec:visual} and Figure \ref{fig:examples}.}
\label{fig:cas}
\end{figure*}

\begin{figure*}
\centering
\includegraphics[scale=0.45]{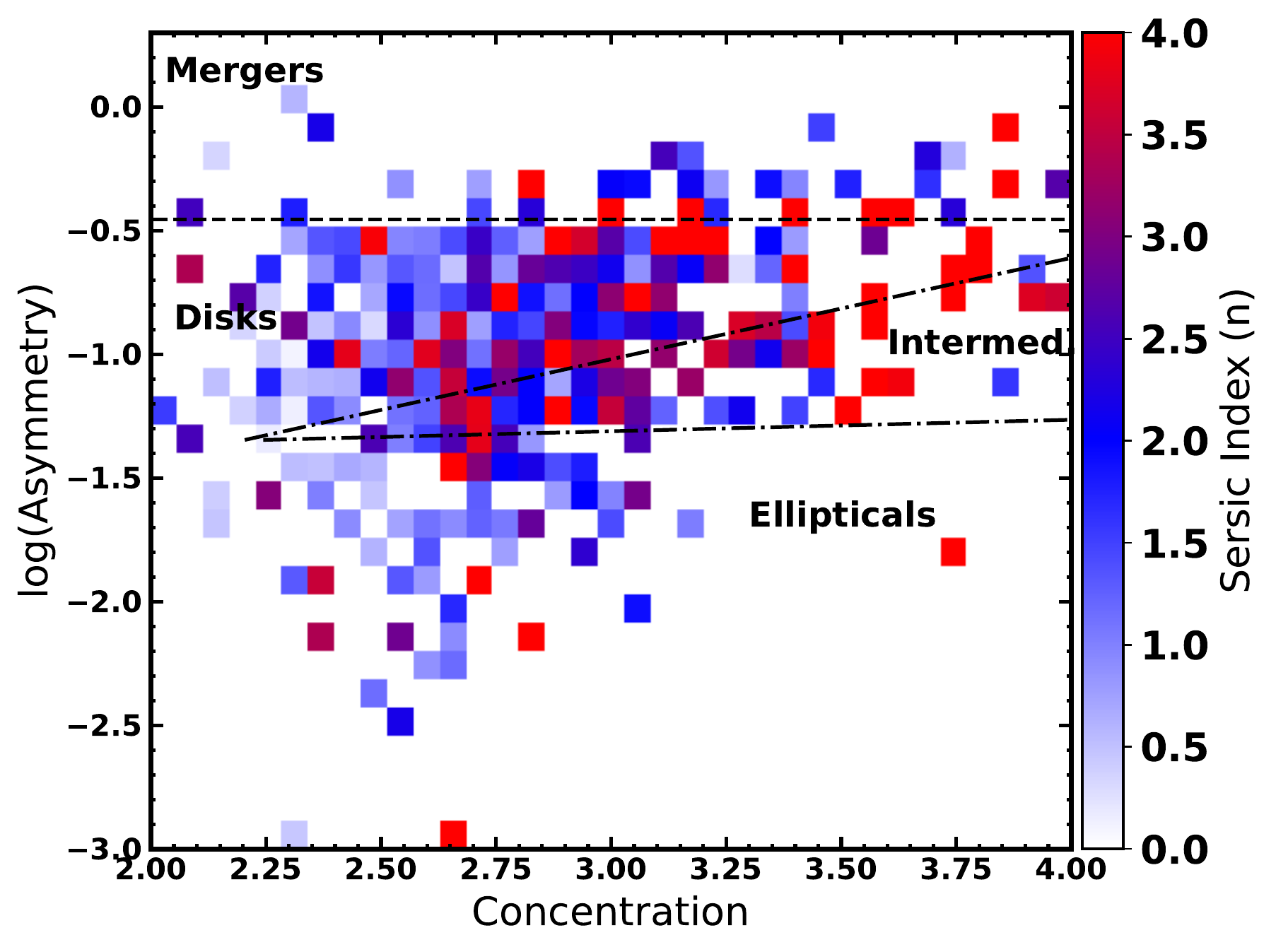}
\includegraphics[scale=0.45]{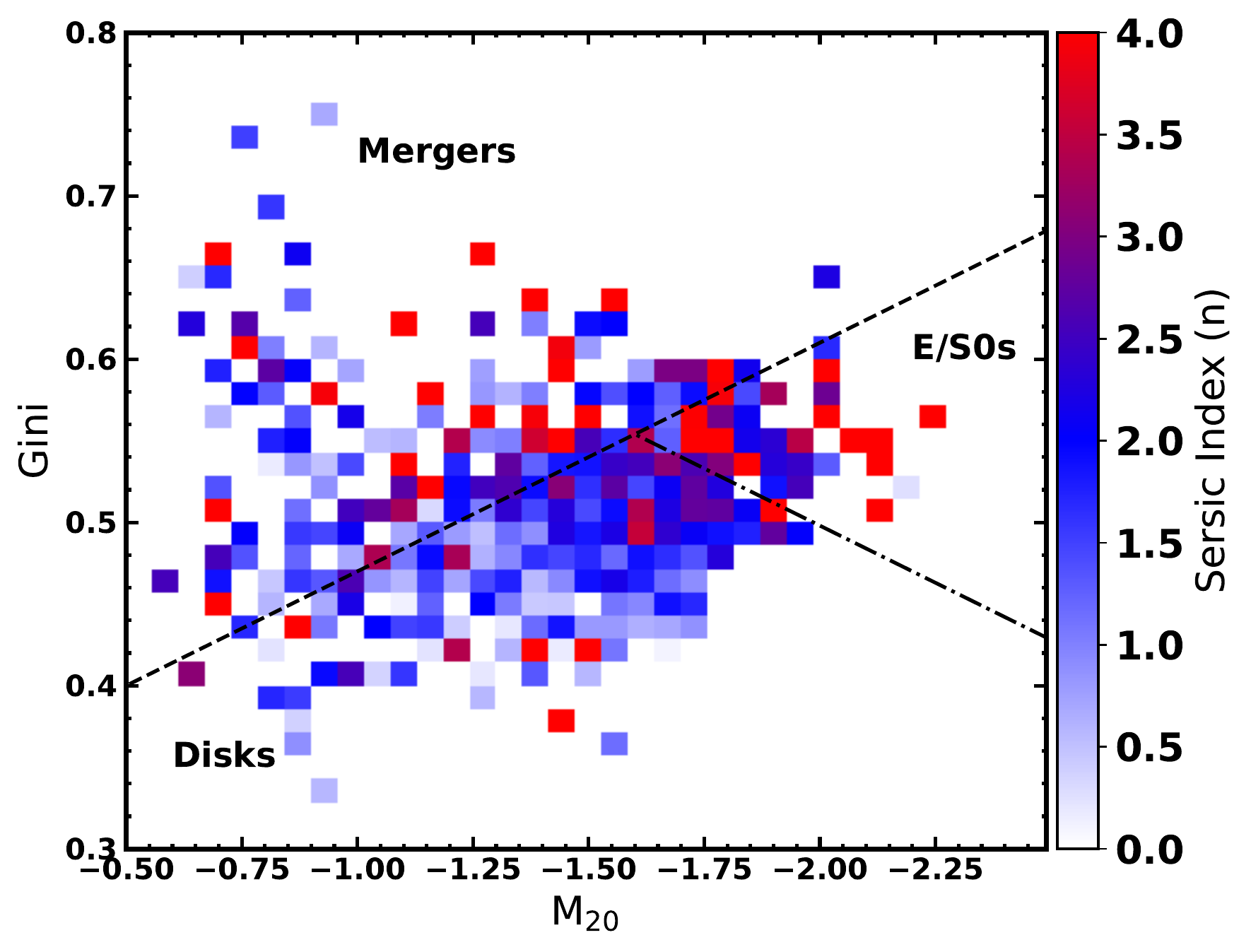}
\caption{The rest-frame optical asymmetry value as a function of concentration (left) and The rest-frame optical Gini value as a function of M$_{20}$ (right) for all galaxies at $z>3$, color coded by the median S\'ersic index in each bin. The dividing lines are the same as described in Figure \ref{fig:cas}}
\label{fig:cas_gm20_sersic}
\end{figure*}

\subsection{Comparison with Non-parametric Measures}\label{sec:statmorph}

As described in Section \ref{sec:statmorph-fits}, we used \texttt{statmorph} to measure non-parametric image statistics for all of the HST-selected $z>3$ galaxies in our sample across all NIRCam filters. We use the same filters corresponding to the rest-frame optical emission for each galaxy as we did for the above parametric comparison. In total, 81\% of the galaxies in our sample have a reliable fit from \texttt{statmorph}; Figures \ref{fig:cas} and \ref{fig:cas_gm20_sersic} highlight two of the commonly used methods to separate galaxies into the standard Hubble types and identify mergers based on these image statistics (e.g., \citealt{ber2000, con2003, lotz2004, lotz2008}).

The top panel of Figure \ref{fig:cas} shows the location of each galaxy on the asymmetry-concentration plane, with the classic lines used to mark the boundaries between disk galaxies, elliptical galaxies, intermediate galaxies, and mergers for moderate redshift HST images \citep{ber2000}. While these boundaries do not cleanly separate $z>3$ galaxies into different types relative to their visual classifications, a few trends can be seen. On average, galaxies with a spheroid have a higher concentration (C) than those with disks and irregulars. Similarly, irregular galaxies have a higher asymmetry value (A), on average. Very few galaxies lie above the classic demarcation for mergers \citep{con2003}, and those that do span the full range of visual morphologies, albeit with a higher fraction of irregulars. Figure \ref{fig:cas_gm20_sersic} shows the distribution of the same galaxies on this plane, but color coded by the median S\'ersic index for each bin. This distribution highlights the correlation of the concentration value with the S\'ersic index for the sample.

The bottom panel of Figure \ref{fig:cas} shows the location of each galaxy on the Gini-M$_{20}$ plane. The lines mark the boundary between disk galaxies, ellipticals/S0s, and mergers based on nearby galaxies and then adjusted for galaxies at $z\sim1$ \citep{lotz2004,lotz2008}. While there is no discernible difference between $z>3$ disks and spheroids with this diagnostic (similar to what has been seen at lower redshift and in simulations, e.g., \citealt{Lotz2008b,Kartaltepe10b,pearson2019a}), irregular galaxies have higher Gini and M$_{20}$ values, on average. 28\% of galaxies with irregular features lie above the merger line, whereas only 17\% of disks and 15\% of spheroids do.  The right-hand panel of Figure \ref{fig:cas_gm20_sersic} shows the distribution of the same galaxies on this plane, but color coded by the median S\'ersic index for each bin. Galaxies that occupy the ellipticals/S0 portion of this plane have a higher S\'ersic index, on average. A significant fraction of the galaxies in the merger region of the plane also have a higher average S\'ersic index.

\section{Discussion}\label{sec:discussion}

\subsection{Disks and Spheroids in the Early Universe}

We find that galaxies detected by both HST and JWST in the $z>3$ universe have a wide diversity of morphologies. Overall, $\sim$60\% of these galaxies have disks (including those with spheroids and/or irregular features as well) at $z=3-4$, and this fraction has an apparent downward trend with increasing redshift. Other early JWST studies have identified candidate disk galaxies at these redshifts (e.g., \citealt{Ferreira2022b, Ferreira2022c, Robertson2022b}) and find similar fractions. Galaxies with spheroids make up $\sim$40\% over this redshift range, with some variations in the higher redshift bins that are likely related to the the small numbers in these bins overall or the difficulty in identifying low surface brightness features at these redshifts. Galaxies with a pure spheroid, i.e., without discernible disks or irregular features, make up $\sim20\%$ across the full redshift range, roughly consistent with the findings of \cite{Ferreira2022b}. While the fraction of all galaxies with irregular features is roughly constant at all redshifts ($\sim$40-50\%), the fraction of galaxies that are purely irregular (i.e., those that have no discernible disk or spheroidal features) increases from $\sim12\%$ at $z=3-3.5$ to $\sim20\%$ at $z>4.5$ This fraction is lower than that reported by \cite{Ferreira2022b,Ferreira2022c}. Slight differences among these early studies likely arise due to the different classification schemes and mass ranges used.

The distribution of axis ratios and sizes presented in Section~\ref{sec:galfit} and Figures \ref{fig:n_Re_Q} and \ref{fig:Q_Re_Redshift} suggest that our $z>3$ sample indeed contains a mix of true disks and spheroids. The distribution of axis ratios for the galaxies classified as Spheroid Only is consistent with that expected from a population of triaxial ellipsoids, while the relatively flat distribution and lower median for the Disk galaxies is expected for a population of disks with exponential profiles and random orientations (e.g., \citealt{Padilla2008}). The axis ratio distribution for the Spheroid Only group peaks at $b/a>0.6$ across the entire redshift range of our sample, while the median for the Disk Only group remains at $b/a\sim0.4$. The Disk+Spheroid and Spheroid+Irregular populations have axis ratio and size distributions that are intermediate.  Previous theoretical work has found that galaxy shapes have evolved over time from prolate to oblate as they transition from having dark matter dominated interiors to baryonic matter interiors following a compaction event (e.g., \citealt{Ceverino15, Tomassetti2016})  and that this transition happens earlier for more massive galaxies (e.g., \citealt{Zhang19a}).

\begin{figure*}
\hspace*{-0.1in}
\includegraphics[scale=0.77]{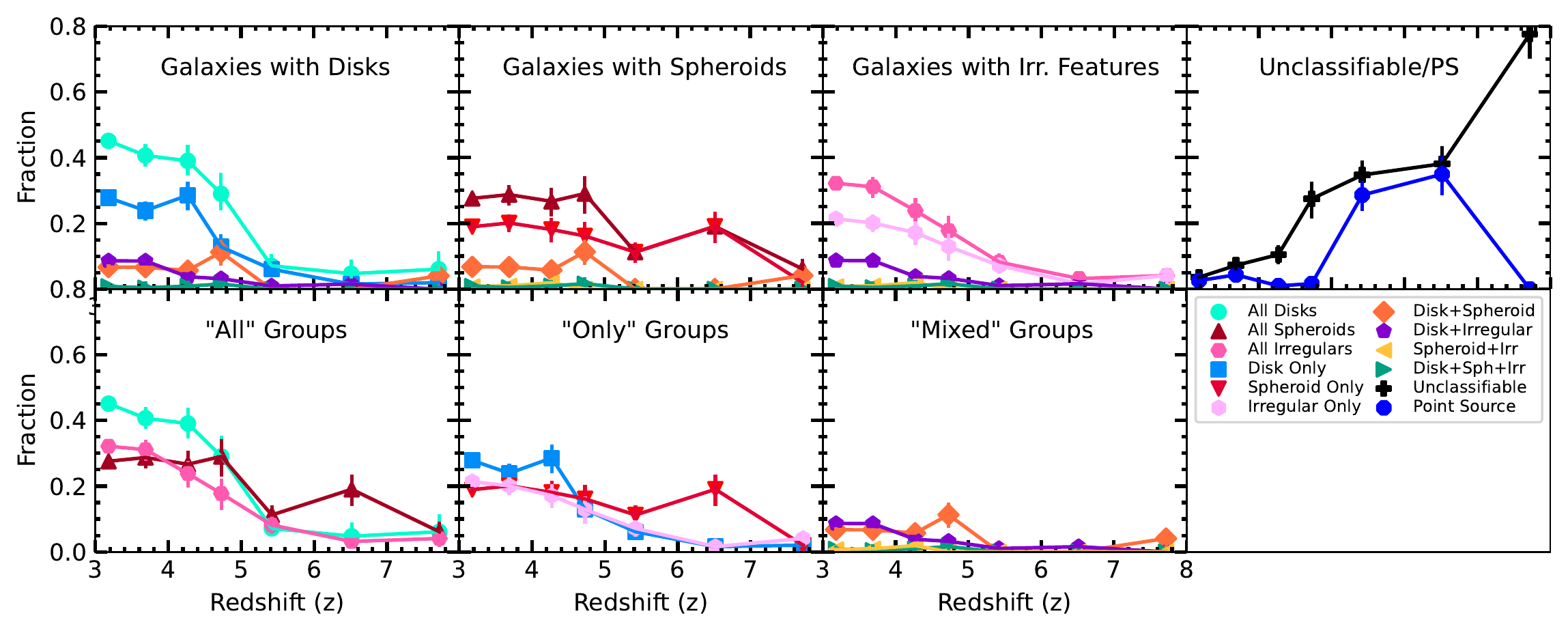}
\caption{The fraction of $z>3$  galaxies detected HST with M$_{\star}>10^{9}$\,M$_{\odot}$ as a function of redshift for each morphology class based on the CANDELS HST visual classifications of \cite{Kartaltepe15a}. The top row, from left to right shows galaxies with disks, galaxies with spheroids, galaxies with irregular features, and Point Sources and Unclassifiable galaxies. The bottom row shows all of the same morphological groups, but divided in different ways for easy comparison. From left to right, the combination of all disks, all spheroids, and all irregulars; the combination of Disk Only, Spheroid Only, and Irregular Only groups; and finally, the remaining mixed groups. Error bars represent the 1$\sigma$ binomial confidence limits given the number of objects in each category, following the method of \cite{cam2011}.}
\label{fig:fractions-hst}
\end{figure*}

\begin{figure*}
\hspace*{-0.1in}
\includegraphics[scale=0.79]{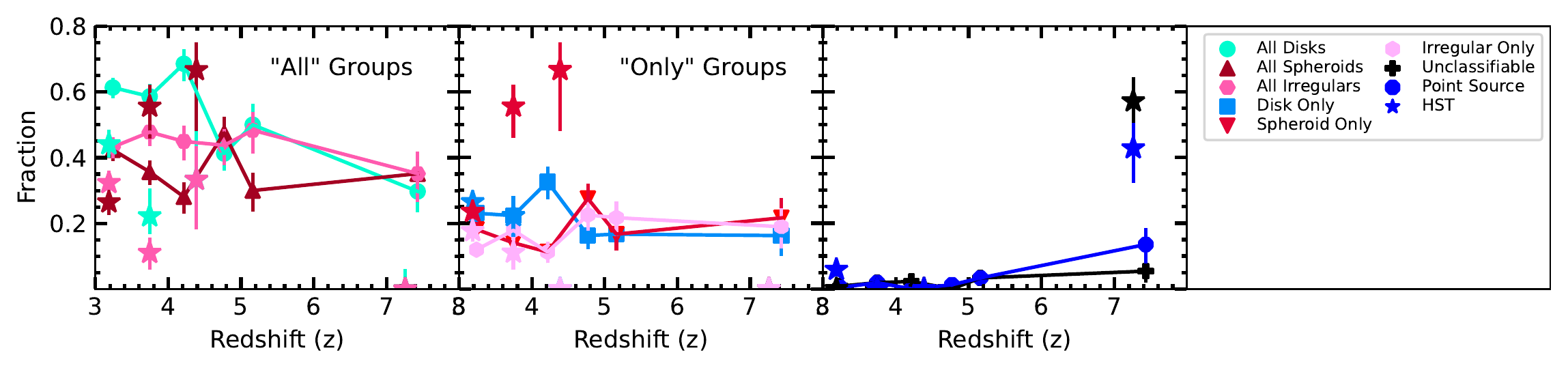}
\caption{The fraction of $z>3$  galaxies detected HST with M$_{\star}>10^{9}$\,M$_{\odot}$ as a function of redshift for each morphology class based on the CEERS JWST visual classifications (as in Fig.~\ref{fig:fractions_z}) compared to the CANDELS HST visual classifications of \cite{Kartaltepe15a} (as in Fig.~\ref{fig:fractions-hst}) for the 59 objects from EGS that were bright enough to be classified.  From left to right, the combination of all disks, all spheroids, and all irregulars; the combination of Disk Only, Spheroid Only, and Irregular Only groups; and finally, objects that were unclassifiable or point sources. Error bars represent the 1$\sigma$ binomial confidence limits given the number of objects in each category, following the method of \cite{cam2011}.}
\label{fig:fractions-hst-same}
\end{figure*}

It is worth noting that selection effects may be partially responsible for the axis ratio distributions observed for these objects. For example, it has been shown that the distribution of axis ratios has a strong dependence on the mass and luminosity of the galaxy population (e.g., \citealt{Padilla2008, Zhang19a}). Galaxy orientation also plays an important role in this distribution, as face-on disks may be more difficult to detect than edge-on disks at the magnitude limit (e.g., \citealt{Elmegreen2005}) and the presence of dust can impact the measured axis ratios \citep{Padilla2008}. The size of the current sample does not allow binning by mass, luminosity, or finer morphology groupings, however, future work with larger sample sizes will allow greater exploration of this parameter space.

To summarize, we see evidence for galaxies with established disks and spheroidal morphologies across the full redshift range of our sample. 
We emphasize that the fractions quoted here are apparent fractions only and that several observational effects likely play a role in these measurements (as discussed in Section~\ref{sec:diff}). Further work is needed to quantify our ability to pick out disk features and resolve spheroidal galaxies in JWST images at varying image depths in order to quantify the true fraction of galaxies with disks and spheroids at these redshifts. Likewise, larger samples, particularly at $z>6$, will be needed to truly establish when the first disks began to form, when disks grew their bulges, and when  spheroids emerged.

\subsection{Comparison between JWST and HST Morphologies} \label{sec:diff}

Figure \ref{fig:fractions-hst} shows the morphological fractions as a function of redshift based on CANDELS HST imaging and using the visual classifications of \cite{Kartaltepe15a} for all $z>3$ galaxies in all five CANDELS fields (1375 galaxies in total). The HST classifications were limited to galaxies with F150W$<24.5$, because fainter galaxies could not be reliably classified, and so only 59 galaxies out of the 850 in our sample are bright enough to make that cut. A comparison of the JWST morphological fractions with those 59 specific galaxies are shown in Figure \ref{fig:fractions-hst-same}. Based on the HST imaging alone, a smaller fraction of galaxies at $z=3.0-4.5$ have disks ($\sim$40\%) and a larger fraction are pure spheroids ($\sim$20\% at $z=3.0-5.0$). The fraction of galaxies that are only irregular is small and drops with redshift, from $\sim$5\% at $z>6$. The fraction of galaxies that are unclassifiable rises sharply, $\sim$5\% at $z=3.5$ to $\sim$35\% at $z=5.5$ to $\sim$80\% at $z>7$. Likewise, $\sim$30\% are unresolved at $z=5-7$.  Among the 59 galaxies with both HST and JWST classifications, a higher fraction is classified as a spheroid and a lower fraction is classified as a disk or irregular with HST than with JWST. At $z>6$, all of the objects were unclassifiable or unresolved with HST.

\begin{figure}
\vspace*{0.2in}
\hspace*{-0.1in}
\includegraphics[scale=0.46]{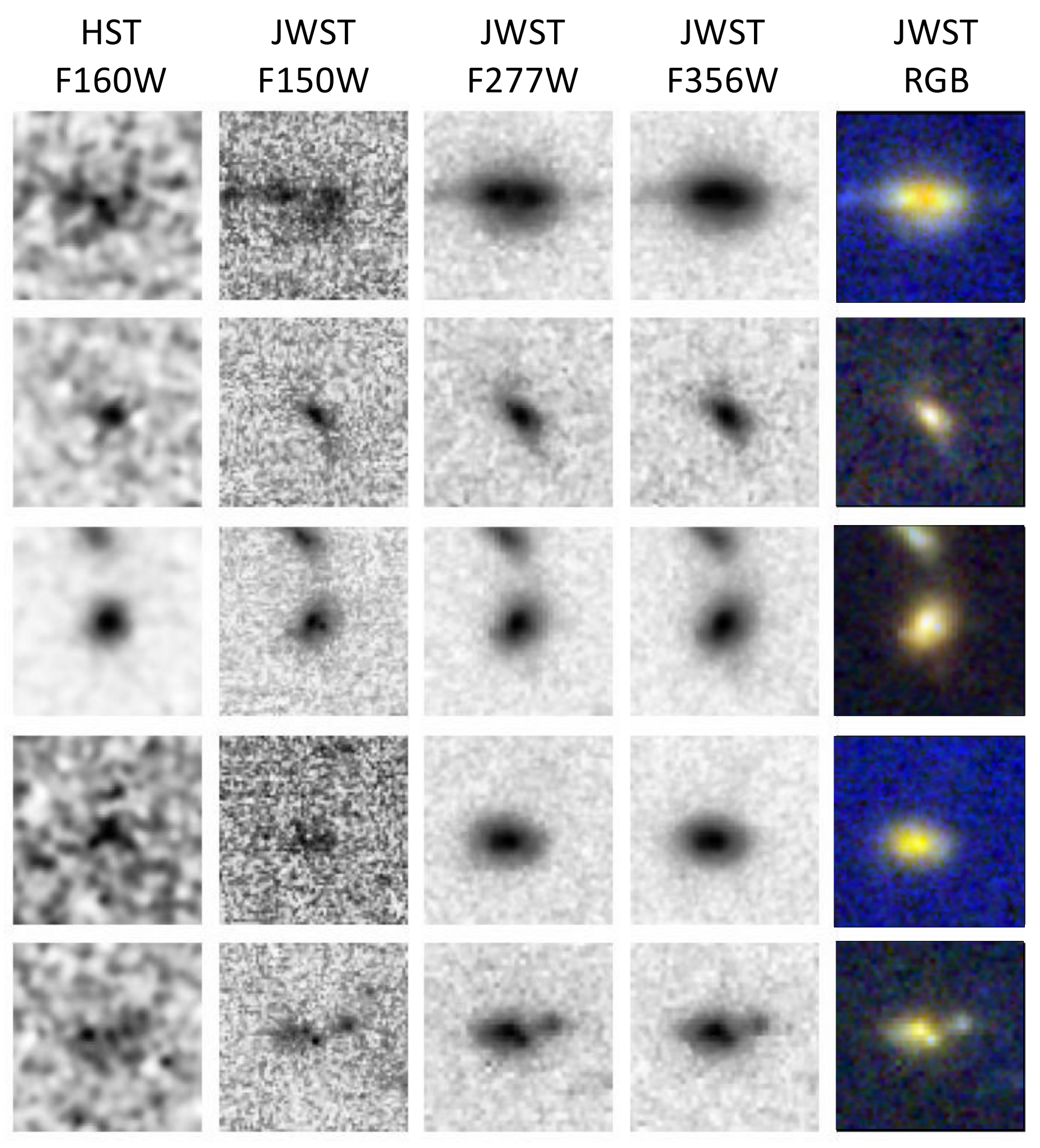}
\caption{HST and JWST postage stamps for five example galaxies with different morphologies in HST F160W images compared to JWST images or differences across the JWST filters. The F150W, F277W, and F356W filters are shown along with an RGB combination of these three filters. Each stamp is 2$''$ on a side. }
\label{fig:diff}
\end{figure}

The large difference seen between the HST and JWST morphologies at these redshifts is expected and is due to the difference in depth and wavelength coverage. 488 galaxies were flagged by at least one classifier as having a different morphology in the JWST images compared to the HST images (159 galaxies were flagged by two out of three classifiers, see Fig.~\ref{fig:diff} for examples). A significant number of galaxies with disks were previously identified as spheroids because of their compact central morphologies, with low surface brightness disk features that only became visible with deeper imaging (see, for example, \citealt{Conselice2011, Mortlock13a, Kartaltepe15a}). This suggests that some fraction of the spheroidal galaxies observed with JWST, particularly those that are faint and/or at higher redshifts, possibly have unobserved disks as well. It is not likely that these disks would previously have been identified as irregular, except for some at the low redshift end, as these irregular features are also too faint to be easily identified at these redshifts with HST. At the low redshift end ($z=3-3.5$), some disks may have been classified as irregular if the the HST data only only picked up the brighter star forming clumps rather than the underlying disk structure. 

To explore the impact of observed wavelength on the classifications, classifiers were also asked to flag objects for which their morphology changes (i.e., they would have selected different main morphology classes) between the NIRCam short wavelength filters and the long wavelength filters. At least one classifier chose this flag for 190 galaxies and two out of three chose it for 37 (see Fig.~\ref{fig:diff} for examples). Note that some of the difference seen across the different filters may be due in part to the increased resolution of the short wavelength bands. This flag was rarely chosen, suggesting that the depth of the JWST images is the primary driver in the morphological differences observed between JWST and HST.

\subsection{Comparison with Expectations from Theory}

\begin{figure*}
\includegraphics[scale=0.7]{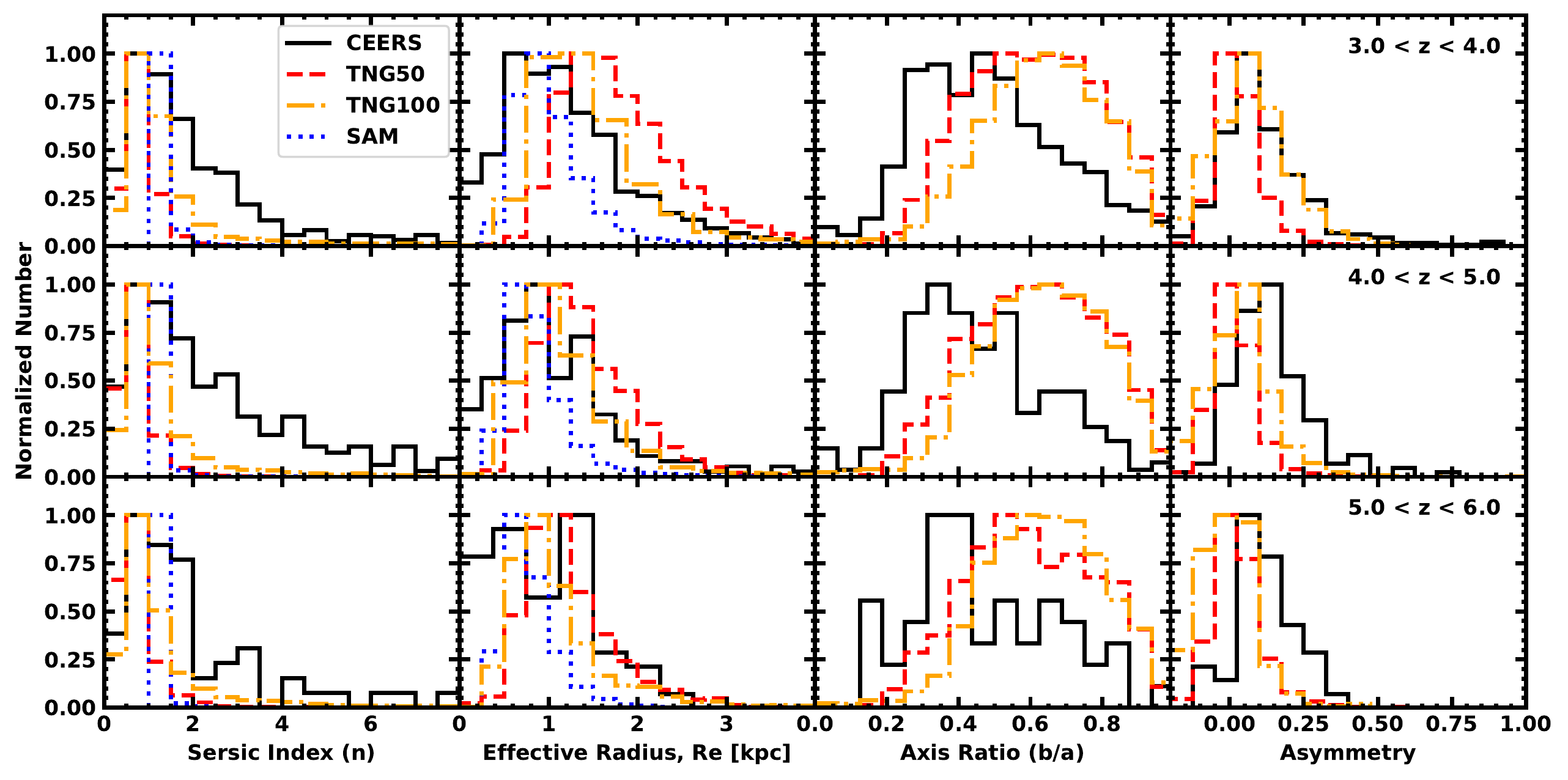}
\caption{The distribution of S\'ersic index, size, axis ratio, and asymmetry of the $z>3$ CEERS galaxy in three different redshift bins compared to the distribution from the CEERS mock catalog derived from the Santa Cruz Semi-analytic model (blue dotted line; \citealt{Somerville2015, Somerville2021, Yung2019, yung2022}), measurements from mock images based on IllustrisTNG50 (red dashed line; \citealt{Costantin22b}), and measurements from mock images based on IllustrisTNG100 (orange dash-dotted line; \citealt{Rose22a}).}

\label{fig:models}
\end{figure*}

We compare the results of our surface brightness profile fits and non-parametric fits with the results from mock images and catalogs based on several different simulations in Figure \ref{fig:models} in three different redshift bins.

First, we use a mock galaxy catalog based on the Santa Cruz Semi-analytic model (SAM) and publicly available as part of the CEERS simulated data release SDR3\footnote{https://ceers.github.io/sdr3.html\#catalogs}. The CEERS mock galaxy catalog is an augmented version of the EGS lightcone presented by \citet{yung2022}, which spans 782 arcmin$^2$ between $0 < z \lesssim 10$ and contains galaxies $-16 \gtrsim M_{\rm UV} \gtrsim -22$. The physical properties of the galaxies are modeled with the physics-based Santa Cruz SAM \citep{Somerville2015, Somerville2021, Yung2019}. The sizes of the disk components of galaxies are computed based on the ansatz that the specific angular momentum of the halo gas is equal to that of the dark matter halo, and that it is conserved during disk formation \citep{Mo:1998,somerville:2008}.  

We also use the publicly available\footnote{https://ceers.github.io/ancillary\_data.html} mock images and derived morphological catalogs of \cite{Costantin22b} and \cite{Rose22a}, which use the IllustrisTNG cosmological simulation\footnote{https://www.tng-project.org/}. The IllustrisTNG project \citep{TNG1Springel2018,TNG2Naiman2018,TNG3Nelson2018,TNG4Pillepich2018,TNG5Marinacci2018} is a series of large cosmological magnetohydrodynamical simulations of galaxy formation and is an update to the original Illustris-1 simulation \citep{vogelsberger14}. It consists of three different runs that span a range of cosmological volumes and resolutions: TNG50, TNG100, and TNG300. 

\cite{Costantin22b} produced synthetic images of TNG50 galaxies and used \texttt{MIRaGe}\footnote{https://github.com/spacetelescope/mirage} (Multi Instrument Ramp Generator) to simulate raw NIRCam images for the CEERS depth and filter combination. These images were then reduced using the JWST pipeline. Morphological measurements were made using \texttt{statmorph}. \cite{Rose22a} produced noiseless synthetic images with TNG100 galaxies using the public visualization API \citep{Nelson2019} in each of the CEERS filters. These images were then convolved with the model PSF for each filter using \texttt{WebbPSF} \citep{perrin2014}. Poisson noise and background noise estimated from the JWST exposure time calculator \citep{pont2016} were added to create mock images at the CEERS depth. Parametric models were fit using \texttt{Galapagos-2} and \texttt{GalfitM} while non-parametric fits were performed using \texttt{statmorph}.

Figure \ref{fig:models} compares the distribution of the S\'ersic indices and sizes of galaxies from the SAM, TNG50, and TNG100 to the distribution measured from CEERS galaxies (Section \ref{sec:galfit} and Figure \ref{fig:galfit}). The overall distributions from the SAM have very similar peaks with a narrower distribution, which holds for all three redshift bins. The S\'ersic index for both TNG50 and TNG100 peak at lower values than the CEERS galaxies and have narrower distributions at all redshifts. At $z=3-4$, TNG50 galaxies have larger sizes than TNG100 galaxies and even larger than both the SAM galaxies and the observed CEERS galaxies. At $z>4$ the distributions match more closely. At all redshifts, the simulations do not contain the smaller (lower R$_{e}$) more compact (larger $n$) galaxies that we observe with JWST CEERS imaging.

Figure \ref{fig:models} also compares the measured axis ratio and asymmetry value for the TNG50 and TNG100 galaxies to the distribution from CEERS. In all three redshift bins, the axis ratios of the TNG50 and TNG100 galaxies match each other well, but peak at higher $b/a$ ($\sim$0.6) and fall off more sharply at lower values than the observed CEERS galaxies. At $z=3-4$, the asymmetry distributions for TNG50, TNG100, and CEERS are well-matched, but the TNG50 and TNG100 distributions shift toward lower (more negative) values at higher redshift. Negative asymmetry values are unphysical and typically result from low $S/N$ sources, where the source is very close to the background level that is being subtracted when making the asymmetry measurement.

Overall, the agreement between our measurements for the $z>3$ JWST CEERS galaxies and the various simulations is encouraging. The differences seen (for example, the difference in axis ratio and the lack of small compact galaxies in the simulations) are worthy of a more in-depth look in order to determine if there are selection effects impacting the results or if there is an actual physical difference between galaxies in these simulations and those in the real observed universe.

\section{Summary and Conclusion}\label{sec:summary}

In this work, we have conducted a comprehensive analysis of 850 $z>3$ galaxies detected in both HST CANDELS imaging of the EGS field and JWST CEERS NIRCam imaging. These galaxies were visually classified by three people each, their parametric morphologies were measured using \texttt{Galfit} and \texttt{Galapagos-2/GalfitM}, and their non-parametric morphologies were measured using \texttt{statmorph}. Our visual classification scheme contains classes that are intentionally not mutually exclusive so that we can track the properties of galaxies with different components separately.  We compare our results to morphology measurements based on the HST imaging alone, as well as several cosmological simulations. Our results are summarized as follows:

\begin{enumerate}

\item Galaxies detected by both HST and JWST in the $z>3$ Universe have a wide diversity of morphologies. Galaxies that have disks make up a large fraction of our sample at all redshifts, from $\sim60\%$ at $z=3-4$ to $\sim30\%$ at $z>6$. Galaxies with spheroids make up $\sim40\%$ across the full redshift range, while pure spheroids without a disk component or irregular features make up $\sim20\%$. The fraction of galaxies with irregular features is roughly constant at all redshifts ($\sim40-50\%$), while those that purely irregular (with no evidence for a disk or spheroidal component) increase from $\sim12\%$ at $z=3.0-3.5$ to $\sim20\%$ at $z>4.5$.

\item Significant differences are seen between JWST morphologies and the HST morphologies for the same galaxies. With only HST imaging, a smaller fraction of galaxies at $z>3$ have disks, spheroid, or irregular features overall due to the larger fraction, particularly at $z>4.5$, that are unresolved or unclassifiable. For resolved classifiable galaxies, the observed difference in classification is largely driven by low-surface brightness disks being too faint to capture in the HST imaging.

\item The distributions of S\'ersic index, size, and axis ratios show significant differences between the different morphological groups, as expected. The spheroid population has a broad distribution of S\'ersic index, and therefore, S\'ersic index cannot be used to cleanly separate disk-dominated from Spheroid-dominated galaxies, as has been shown previously based on HST imaging. Galaxies with a spheroid tend to be smaller, on average, than galaxies with disks or irregular features.

\item The distribution of axis ratios for the Spheroid Only galaxies peaks at high values and is consistent with a triaxial population. The Disk Only, Irregular Only, and Disk+Irregular galaxies peak at lower values with an overall broad distribution, while the Disk+Spheroid and Spheroid+Irregular groups are intermediate. In general, smaller galaxies tend to be rounder.

\item While classical classification boundaries using non-parametric measures such as concentration, asymmetry, Gini, and M$_{20}$ do not cleanly separate galaxies by their morphological type, galaxies with a spheroid have a higher concentration, on average, than disks and irregulars, while irregular galaxies have a higher mean asymmetry value. Irregular galaxies also have higher Gini and M$_{20}$ values on average and are slightly more likely than disks or spheroids to lie above the merger selection line.

\item The distribution of S\'ersic index, size, axis ratio, and asymmetry of the $z>3$ sample is overall well-matched by the distributions from the CEERS mock catalog derived from the Santa Cruz Semi-analytic model, and measurements from mock images based on IllustrisTNG50 and IllustrisTNG100 galaxies. The simulations do not have the small compact galaxies that we observe in CEERS. The axis ratio distribution for TNG50 and TNG100 galaxies peaks at higher $b/a$ and drops off more sharply at lower values than the CEERS galaxies.
\end{enumerate}

Overall, these trends suggest that galaxies with established disks and spheroidal morphologies exist across the full redshift range of this study. Future work with larger samples that capture many more galaxies at the high redshift end in conjunction with observations that can probe their dynamical nature are needed to fully explore the parameter space, understand how these disks and spheroids compare to today's, and quantify the emergence of the first disks and spheroids.

\begin{acknowledgments}

 Support for this work was provided by NASA through grant JWST-ERS-01345.015-A and HST-AR-15802.001-A awarded by the Space Telescope Science Institute, which is operated by the Association of Universities for Research in Astronomy, Inc., under NASA contract NAS 5-26555. This research is based in part on observations made with the NASA/ESA Hubble Space Telescope obtained from the Space Telescope Science Institute, which is operated by the Association of Universities for Research in Astronomy, Inc., under NASA contract NAS 5–26555. JSK would like to acknowledge the important contributions of her cats, T'Pol and Shran, who diligently attended every telecon and assisted in paper editing.

 MHC acknowledges financial support from the State Research Agency (AEI\-MCINN) of the Spanish Ministry of Science and Innovation under the grant and ``Galaxy Evolution with Artificial Intelligence" with reference PGC2018-100852-A-I00, from the ACIISI, Consejer\'{i}a de Econom\'{i}a, Conocimiento y Empleo del Gobierno de Canarias and the European Regional Development Fund (ERDF) under grant with reference PROID2020010057, and from IAC project P/301802, financed by the Ministry of Science and Innovation, through the State Budget and by the Canary Islands Department of Economy, Knowledge and Employment, through the Regional Budget of the Autonomous Community

The authors acknowledge Research Computing at the Rochester Institute of Technology for providing computational resources and support that have contributed to the research results reported in this publication. \href{https://doi.org/10.34788/0S3G-QD15}{https://doi.org/10.34788/0S3G-QD15}. The authors also acknowledge the Texas Advanced Computing Center (TACC) at The University of Texas at Austin for providing HPC resources that have contributed to the research results reported within this paper. \href{http://www.tacc.utexas.edu}{http://www.tacc.utexas.edu}

Some of the data presented in this paper are available on the Mikulski Archive for Space Telescopes (MAST) at the Space Telescope Science Institute. The specific observations can be accessed via \dataset[doi:10.17909/qhb4-fy92]{https://doi.org/10.17909/qhb4-fy92} and \dataset[doi:10.17909/T94S3X]{https://doi.org/10.17909/T94S3X}.

\end{acknowledgments}

\facilities{JWST, HST}

\software{\texttt{Astropy} \citep{astropy},
          \texttt{Drizzle} \citep{fruchter02},
          \texttt{Galapagos-2} \citep{hau2013},
          \texttt{Galfit} \citep{peng2010},
          \texttt{GalfitM} \citep{peng2010, hau2013},
          \texttt{Source Extractor} \citep{ber1996},
          \texttt{Statmorph} \citep{Rodriguez-Gomez19a},
}

\bibliographystyle{aasjournal}

\appendix
\twocolumngrid

\section{Additional Details}  \label{appendix}

\begin{figure}
\includegraphics[scale=0.5]{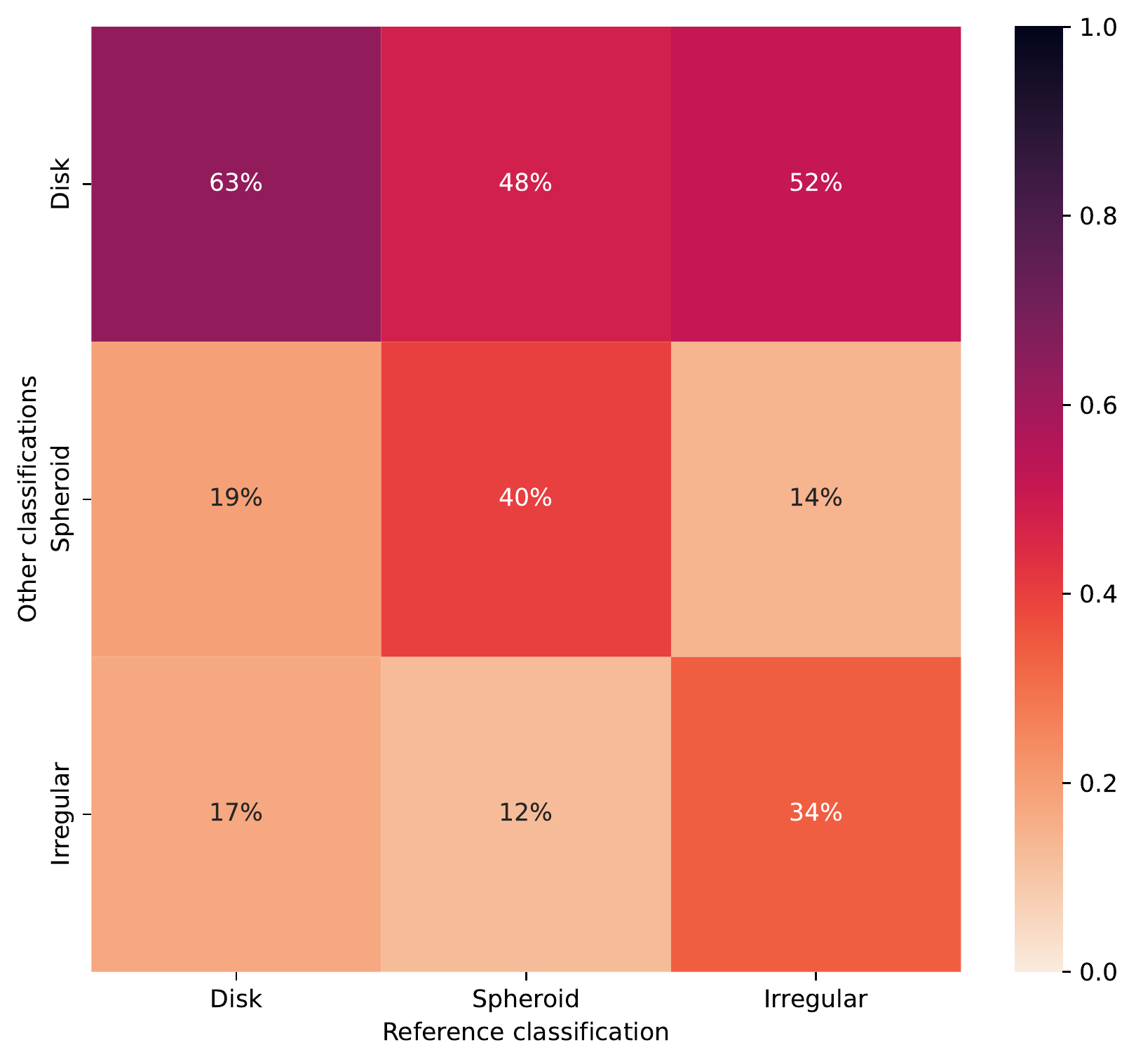}
\caption{A matrix highlighting the level of agreement between the three classifiers for all galaxies with a disk, with a spheroid, or those with irregular features. Overall, classifiers regularly agree when a galaxy has a disk, agree less often about galaxies with a spheroid, and are more likely to disagree about irregular features.}
\label{fig:matrix}
\end{figure}

\begin{figure*}
\includegraphics[scale=0.8]{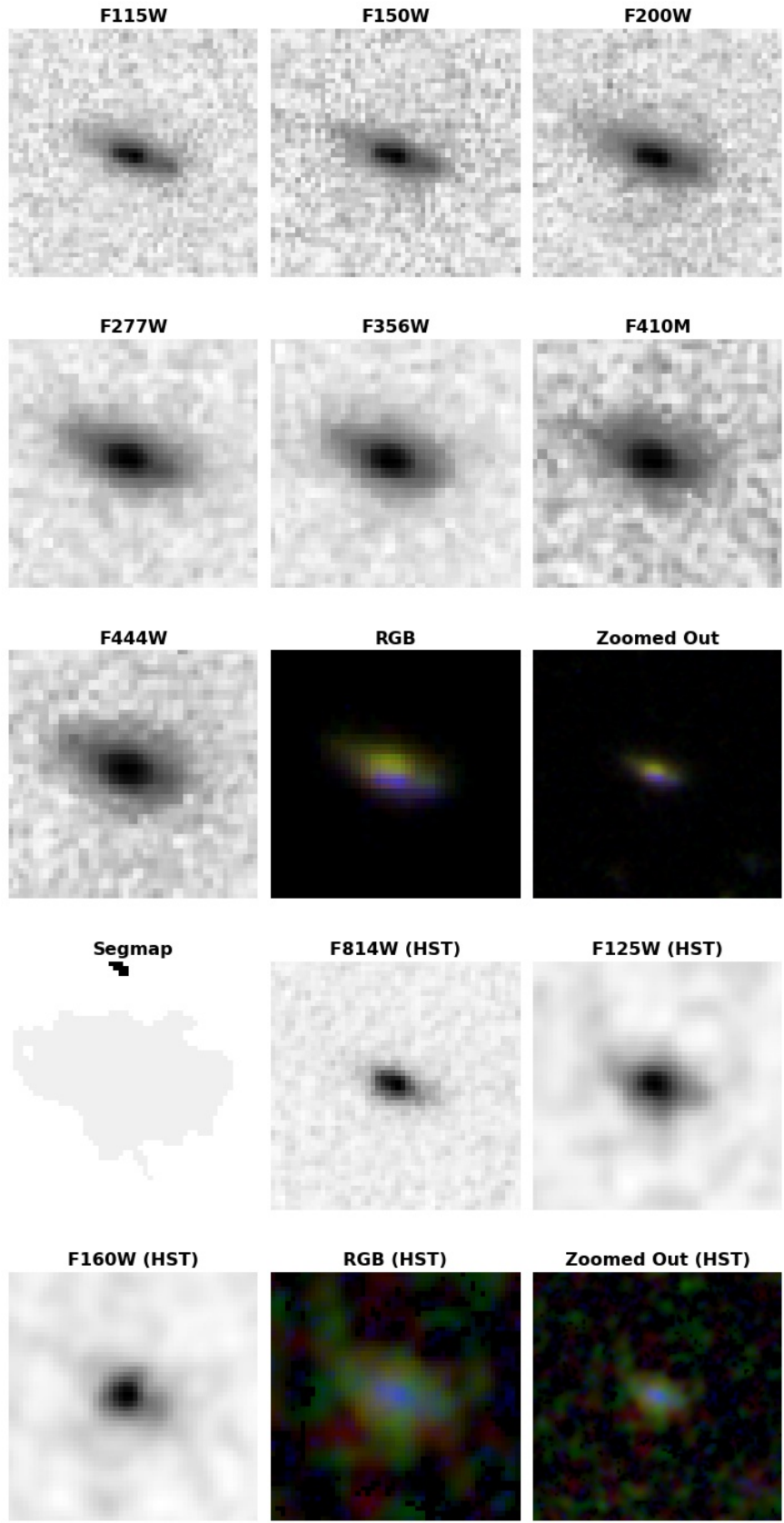}
\caption{An example set of postage stamps for one galaxy (CANDELS ID 16438 with $z=3.8$ and log(M$_{\star}$/M$_{\odot}$)$=10.0$), used for the visual classification of galaxies at $z>3$. The stamps are scaled by the size of the galaxy as measured by \texttt{Source Extractor}, following Equations 2 and 3 of \cite{Haussler2007}, with a minimum size of 100$\times$100 pixels. Two of the three classifiers classified this galaxy as having both a disk and a spheroid, while the third classified it as having a disk only.}
\label{stamps}
\end{figure*}

Here we include some details and additional figures for the visual classifications and \texttt{Galfit} measurements.
Figure \ref{fig:matrix} highlights the level of agreement among the three classifiers for the three options in the main morphological class: disk, spheroid, and irregular. For the visual classifications, an example set of stamps that was shown to the classifiers for one of the galaxies is shown in Figure \ref{stamps}. 

\subsection{\texttt{Galfit} Fits}

We used \texttt{Galfit} to compute parametric fits on the F150W and F200W images. As initial guesses, we use the source location, magnitude, size, position angle, and axis ratios from the \texttt{Source Extractor} catalogs and segmentation maps. We use cutouts of the error array (ERR extension) produced by the JWST pipeline as the input sigma images for each source, which includes Poisson noise from the sources themselves, as well as the usual instrument noise. As input PSFs, we create empirical PSFs for each filter from stacked stars from all four CEERS pointings. For each galaxy, the Kron radius measured by \texttt{Source Extractor} was used to scale the size of the cutout used as input to \texttt{Galfit}. All galaxies in the cutout within three magnitudes of the primary source were simultaneously fit, down to a magnitude limit of 27, with all other sources masked.  Based on our testing, we find that sources fainter than this limit were not reliably fit. We assign each fit a quality flag. A flag of 0 indicates a good fit; a flag of 1 indicates that the fit is suspect, meaning the resulting \texttt{Galfit} magnitude differs substantially from the input \texttt{Source Extractor} magnitudes; a flag of 2 indicates a poor fit, where one or more parameters reached a constraint limit; a flag of 3 indicates thay the fit failed to find a solution; and a flag of 4 indicates that the source was not fit either because it was either an artifact or located too close to the edge. 

Of the 850 galaxies in the $z>3$ sample, 74\% have a \texttt{Galfit} flag of 0, 13\% have a flag of 1, 8\% have a flag of 2, 2\% have a flag of 3, and none have a flag of 4 (these would already have been removed by our initial sample selection). For the comparisons discussed below, we use all of the galaxies with a flag of 0 or 1, representing 87\% of the total sample.

As a check on the quality of our fits, and for consistency, we compared the \texttt{Galfit} and \texttt{GalfitM} fits for the F200W filter and visually inspected the model and residuals. Overall, we see a high level of agreement with no significant offsets for the S\'ersic index, the size, the axis ratio, and the magnitudes between the two measurements for the sources that do not reach a constraint limit. Throughout the paper, we use the \texttt{GalfitM} measurements for the filter closest to the rest-frame optical at the redshift of the galaxy. The median and 16th percentile and 84th percentile range for the S\'ersic index ($n$), the effective radius ($R_e$), and the axis ratio ($b/a$) for each morphological type are given in Table \ref{tab:galfit} and are used throughout the text.

\begin{deluxetable*}{lcccccc}
\tabletypesize{\small}
\tablecaption{Median Properties for Each Morphology Class}
\tablewidth{\textwidth}
\tablehead{
\colhead{Morphology Group} & \colhead{\#} & \colhead{Redshift} & \colhead{log(Mass [M$_{\odot}$])} & \colhead{$n$\tablenotemark{a}} & \colhead{$R_e$ [kpc]\tablenotemark{a}} & \colhead{$b/a$\tablenotemark{a}} 
}
\startdata
All Disks                & 467 & 3.84$\pm$0.87 & 9.31$\pm$0.63     & 1.41$^{+1.38}_{-0.65}$ & 1.19$^{+0.75}_{-0.51}$ & 0.40$^{+0.22}_{-0.14}$ \\
Disks Only               & 192 & 3.86$\pm$0.90 & 9.28$\pm$0.47     & 1.16$^{+0.88}_{-0.40}$ & 1.19$^{+0.61}_{-0.48}$ & 0.40$^{+0.21}_{-0.13}$\\
Disk$+$Spheroids           & 88  & 3.86$\pm$0.98 & 9.29$\pm$1.12   & 2.38$^{+1.06}_{-0.98}$ & 0.83$^{+0.73}_{-0.33}$ & 0.51$^{+0.17}_{-0.16}$\\
Disk$+$Irregulars          & 155 & 3.82$\pm$0.82 & 9.32$\pm$0.41   & 1.12$^{+1.10}_{-0.51}$ & 1.33$^{+0.78}_{-0.42}$ & 0.36$^{+0.21}_{-0.11}$\\
Disk$+$Spheroid$+$Irregulars & 32  & 3.81$\pm$0.60 & 9.49$\pm$0.48 & 2.63$^{+1.63}_{-1.14}$ & 1.33$^{+1.37}_{-0.45}$ & 0.39$^{+0.18}_{-0.10}$\\
All Spheroids            & 323 & 3.94$\pm$1.04 & 9.30$\pm$0.71     & 2.48$^{+2.12}_{-1.34}$ & 0.72$^{+0.66}_{-0.36}$ & 0.56$^{+0.21}_{-0.20}$\\
Spheroid Only            & 156 & 4.03$\pm$1.11 & 9.29$\pm$0.47     & 2.46$^{+2.32}_{-1.41}$ & 0.54$^{+0.42}_{-0.32}$ & 0.64$^{+0.17}_{-0.18}$\\
Spheroid$+$Irregulars      & 47  & 3.90$\pm$1.12 & 9.21$\pm$0.39   & 2.92$^{+2.30}_{-1.96}$ & 0.72$^{+0.81}_{-0.34}$ & 0.55$^{+0.18}_{-0.20}$\\
All Irregulars           & 376 & 3.90$\pm$0.93 & 9.29$\pm$0.43     & 1.37$^{+1.73}_{-0.80}$ & 1.28$^{+0.84}_{-0.57}$ & 0.40$^{+0.21}_{-0.13}$ \\
Irregular Only           & 142 & 4.02$\pm$1.02 & 9.24$\pm$0.44     & 1.12$^{+1.51}_{-0.82}$ & 1.32$^{+0.79}_{-0.57}$ & 0.42$^{+0.17}_{-0.13}$\\
Point Source $/$ Unresolved & 16 & 5.27$\pm$1.82 & 9.66$\pm$0.65   & \nodata & \nodata & \nodata \\
Unclassifiable            & 18 & 4.57$\pm$1.54 & 9.47$\pm$0.84 & \nodata & \nodata & \nodata \\
\enddata
\tablenotetext{a}{Sersic index ($n$), effective radius (R$_{e}$ in kpc), and axis ratio ($b/a$) as measured in the NIRCam filter that most closely represents the rest-frame optical for the redshift of the galaxy: F277W for $3.0<z<4.0$, F356W for $4.0<z<4.5$, and F444W for $z>4.5$.} 
\label{tab:galfit}
\end{deluxetable*}

\end{document}